\documentclass[twocolumn]{aastex62} 
\usepackage{hyperref,enumerate}
\input{colordvi}
\listofchanges
\hypersetup{
    unicode=false,                  % non-Latin characters in Acrobat book
    pdftoolbar=true,                % show Acrobat toolbar?
    pdfmenubar=true,                % show Acrobat menu?
    pdffitwindow=true,              % window fit to page when opened
    pdfstartview={FitH},            % fits the width of the page t
    pdftitle={R443130},             % title
    pdfauthor={vplacco},            % author
    pdfsubject={Astronomy},         % subject of the document
    pdfcreator={dvipdf},            % creator of the document
    pdfproducer={dvipdf},           % producer of the document
    pdfkeywords={metal-poor stars}, % list of keywords
    pdfnewwindow=true,              % links in new window
    colorlinks=true,                % false: boxed links; true: colored
    linkcolor=red,                  % color of internal links
    citecolor=blue,                 % color of links to bibliography
    filecolor=magenta,              % color of file links
    urlcolor=cyan,                  % color of external links
    breaklinks=true,
    linktocpage
}

\newcommand{\rapo}{r_\ensuremath{\mathrm{apo}}}
\newcommand{\rperi}{r_\ensuremath{\mathrm{peri}}}
\newcommand{\zmax}{Z_\ensuremath{\mathrm{max}}}
\def\vector#1{\mbox{\boldmath $#1$}}

\newcommand{\Teff}{$T_{\rm eff}$}  % Command for Teff, text mode

\newcommand{\abund}[2]{\ensuremath{[\mathrm{#1}/\mathrm{#2}]}}

\newcommand{\metal}{\abund{Fe}{H}}

\newcommand{\teff}{\ensuremath{T_\mathrm{eff}}}
\newcommand{\logg}{\ensuremath{\log\,g}}

\newcommand{\Gyr}{\ensuremath{\mathrm{Gyr}}}

\shorttitle{Discovery of two Bright Carbon-Enhanced Metal-Poor Stars}
\shortauthors{Mardini et al.}
\begin{document}

\title{Metal-poor Stars Observed with the Automated Planet Finder Telescope.
II. Chemodynamical Analysis of Six Low-Metallicity Stars in the Halo System of
the Milky Way}

\author[0000-0001-9178-3992]{Mohammad K.\ Mardini}
\affiliation{Key Lab of Optical Astronomy, National Astronomical Observatories, Chinese Academy of Sciences, Beijing, China}
\affiliation{School of Astronomy and Space Science, University of Chinese Academy of Sciences,  Beijing, China}	
\author[0000-0003-4479-1265]{Vinicius M.\ Placco}
\affiliation{Department of Physics, University of Notre Dame, Notre Dame, IN 46556, USA}
\affiliation{JINA Center for the Evolution of the Elements, USA}
\author{Ali Taani}
\affiliation{Physics Department, Faculty of Science, Al-Balqa Applied University, Al-Salt, Jordan}
\author[0000-0002-0389-9264]{Haining Li}
\affiliation{Key Lab of Optical Astronomy, National Astronomical Observatories, Chinese Academy of Sciences, Beijing, China}
\author[0000-0002-8980-945X]{Gang Zhao}
\affiliation{Key Lab of Optical Astronomy, National Astronomical Observatories, Chinese Academy of Sciences, Beijing, China}
\affiliation{School of Astronomy and Space Science, University of Chinese Academy of Sciences, Beijing, China}	

\correspondingauthor{\\Haining Li [lhn@nao.cas.cn]\\Gang Zhao [gzhao@nao.cas.cn]}

\nocollaboration

\begin{abstract}
In this work, we study the chemical compositions and kinematic properties of six
metal-poor stars with [Fe/H] $< -2.5$ in the Galactic halo. From high-
resolution (R $\sim$~110,000) spectroscopic observations obtained with the Lick/APF, we
determined individual abundances for up to 23 elements, to quantitatively
evaluate our sample. We identify two carbon-enhanced metal-poor stars
(J1630+0953 and J2216+0246) without enhancement in neutron-capture elements (CEMP-no
stars), while the rest of our sample stars are carbon-intermediate. By comparing the
light-element abundances of the CEMP stars with predicted yields from
non-rotating zero-metallicity massive-star models, we find that possible the progenitors
of J1630+0953 and J2216+0246 could be in the 13-25 M$_{\odot}$ mass range, with
explosion energies 0.3-1.8$ \times 10^{51}$ erg. In addition, the detectable
abundance ratios of light and heavy elements suggest that our sample stars are
likely formed from a well-mixed gas cloud, which is consistent with previous
studies. We also present a kinematic analysis, which suggests that most of our
program stars likely \textbf{belong to} the inner-halo population, with orbits
passing as close as $\sim$ 2.9 kpc from the Galactic center. We discuss the
implications of these results on the critical constraints on the origin and
evolution of CEMP stars, as well as the nature of the Population III progenitors
of the lowest metallicity stars in our Galaxy.  
\end{abstract} 

\keywords{Galaxy: halo---techniques: spectroscopy--- stars:fundamental
parameters--- stars: abundances---stars: atmospheres---stars: kinematics and
dynamics --- stars: Population II--- stars: chemically peculiar}

\section{Introduction} 
\label{sec:intro}

The detailed study of the chemical composition of metal-poor stars in the Milky
Way greatly contributes to our understanding of Galactic chemical evolution
(GCE).  The best candidates for such studies are the very metal-poor
(\abund{Fe}{H} $< -2.0$, hereafter VMP), extremely metal-poor
(\abund{Fe}{H} $< -3.0$, hereafter EMP), and  ultra metal-poor
(\abund{Fe}{H} $< -4.0$, hereafter UMP) stars. These second-generation objects
(Pop II) have formed from low-metallicity gas clouds at redshift $\gtrsim 6$
\citep[see][and references therein]{2018ARNPS..68..237F}. Each element in their
atmosphere can potentially help us understand the underlying physical processes
by which our Galaxy evolved chemically, considering that each element might
follow different nucleosynthesis pathway(s). 

A number of previous observational studies have indicated that carbon is
omnipresent in the early universe \citep{2005ARA&A..43..531B,
2007ApJ...655..492A, 2016A&A...588A..37H, 2016ApJ...829L..24P,
2018MNRAS.475.4781C, 2018A&A...614A..68C}. Hence, the discovery and analysis of
carbon-enhanced metal-poor stars ([C/Fe] $\geqslant 0.7$\footnote{Using
\citet{2007ApJ...655..492A}, as the criterion for carbon enhancement.}
hereafter CEMP), suggest that this substantial enhancement could be closely
linked to their formation.  In addition to carbon enhancement, different
abundance ratios of neutron-capture elements are often used to distinguish the
unique nature of CEMP stars:  CEMP-s ([C/Fe] $\geqslant +0.7$, [Ba/Fe] $>
+1.0$, and [Ba/Eu] $> +0.5$),  CEMP-r/s ([C/Fe] $\geqslant +0.7$ and $0.0 <$
[Ba/Eu] $< +0.5$),  CEMP-r ([C/Fe] $\geqslant +0.7$ and [Eu/Fe] $> +1.0$), and
CEMP-no ([C/Fe] $\geqslant +0.7$ and [Ba/Fe] $< 0.0$).

The CEMP-s and CEMP-no subclasses represent the predominant populations of the
CEMP stars \citep{2005ARA&A..43..531B, 2007ApJ...655..492A,
2016A&A...588A..37H,  2016ApJ...833...20Y}. The chemical patterns associated
with CEMP-s stars (high enhancement in carbon and s-process elements) can arise
from an intrinsic (self-enrichment) or an extrinsic (mass transfer from now
white dwarf companion) process. Nevertheless, the overabundance of s-process
elements (e.g., [Ba/Fe] $> +1.0$) support the matter accretion from an
asymptotic giant branch (AGB) companion. These s-process elements
(e.g., Sr, Ba, and Ce) are believed to be synthesized in low- to
intermediate-mass stars (1 to 3 M$_{\odot}$), with low neutron densities ($n_{n}
\approx 10^{6} $ cm $^{-3} - 10^{10}$ cm $^{-3}$) and $^{13}$C($\alpha,
n$)$^{16}$O as the main source of neutrons, which are eventually
transferred and mixed into the atmosphere of the long-lived companion
\citep[observed as a CEMP-s star,][]{2005ApJ...625..825L,
2014MNRAS.441.1217S}. 
 
CEMP-r/s and CEMP-r are much less frequent when compared with other subclasses.
The origin of their enhancement (in r-process elements) is still widely
debated. In contrast to the s-process elements, a high neutron density ($n_{n}
> 10^{22}$ cm $^{-3}$) is the key to synthesize these
unstable neutron-rich isotopes (e.g., Eu, Os, and Ir). In the past, there have
been many astrophysical sites proposed for r-process production. Still a few
can provide such high neutron density environment \citep[see][and references
therein]{2017ARNPS..67..253T}. Presently, the possible sites are confined to
magnetorotationally jet-driven supernovae (SN), core-collapse SN, neutron
stars mergers, and neutron star-black hole mergers
\citep[e.g.,][]{2018ARNPS..68..237F}.
 
The CEMP-no stars, which are believed to be the direct descendants of objects
formed shortly after the big bang (Pop III), \citep{2013ApJ...773...33I,
2014ApJ...790...34P,2016A&A...586A.160H}, dominate the lowest-metallicity regime
(e.g., \citealt{2002Natur.419..904C, 2005Natur.434..871F, 2011Natur.477...67C,
2014Natur.506..463K, 2018MNRAS.481.3838S}) and reside in the main-sequence,
subgiants, or red giant phase.  Their evolutionary stages and chemical patterns
(excess in carbon with low abundances or absence of neutron-capture elements)
suggest that a binary companion or self-enrichment are unlikely the sources of
their chemical patterns. Therefore, a distinct enrichment channel may have
taken place (e.g., \citealt{2014MNRAS.441.1217S, 2016A&A...588A...3H}).
\textbf{A spinstar is one possible candidate; these rapidly rotating
massive ultra metal-poor (\abund{Fe}{H} $< -6.0$) stars can produce large
amounts of carbon \citep{2006A&A...447..623M, 2007A&A...461..571H,
2012A&A...538L...2F, 2015A&A...576A..56M}.  Another proposed scenario for the
carbon enhancement is pollution from faint supernovae associated with Pop III
stars, with mixing-and-fallback \citep{2003Natur.422..871U,
2007ApJ...660..516T, 2010ApJ...724..341H, 2013ARA&A..51..457N,
2014ApJ...785...98T,2019arXiv190403211E}. This faint SN ejects less iron and
thus increases the [C/Fe] ratio, as they do not have enough energy to eject all
its material into its surroundings. Therefore, only the outer layers with the
lightest elements are ejected while the inner part falls back onto the neutron
star or black hole.}  At present, none of the above scenarios can explain the
full chemical patterns that have been observed in CEMP-no stars. \textbf{In
general, the unique chemical patterns observed in the sub-classes of CEMP stars
result from the differences in the astrophysical sites responsible for the
nucleosynthesis products they now mixed in their atmospheres.}

The Milky Way galaxy has three main components (with respect to its visible
matter): the bulge (a very luminous and dense structure, hosting ongoing star
formation and the Galactic supermassive black hole), the disk (a flattened
region surrounding the bulge, hosting young stars), and the halo (an extend
structure, primarily containing old field stars). \textbf{Metal-poor stars
reside primarily in the Galactic halo system, which} is believed to have at
least two extended regions (the inner-halo and the outer-halo), with different
metallicity distributions, kinematics, and spatial density profiles \citep[e.g.,
][]{2012AAS...21922206B}. The halo system also hosts several stellar streams,
overdensities \citep{2009ApJ...693.1118G}, and a recently discovered large
structure in the inner region, product of a past merger event
\citep{2018Natur.563...85H}.  \textbf{The outer halo exhibits a fraction of CEMP
stars twice as large as the inner halo in the metallicity interval
$-2.5<$[Fe/H]$-2.0$.  Such an increase in frequency of CEMP stars can be
explained as a population-driven effect, due to the fact that the outer halo is
the dominant component at large distances from the galactic plane and at
metallicities [Fe/H]$<-2.0$.}
\textbf{More recently, many studies have suggested that the relative numbers of
CEMP-no stars compared to CEMP-s stars are found to vary between the inner- and
outer-halo \citep{2014ApJ...788..180C, 2017ApJ...835...81B, 2018ApJ...861..146Y,
Mardini_2019}. Based on the stellar distances, those authors and many others
suggested that the frequency of the CEMP-no stars is higher in the outer halo,
while the frequency of the CEMP-s stars is higher in the inner halo. } 

In this paper, we report on the discovery of six stars with \abund{Fe}{H} $<
-$2.5  selected from the LAMOST database, including two CEMP stars. This is a
continuation of the work presented in \citet[][hereafter Paper
I]{Mardini_2019}. By analyzing the chemical abundances and  kinematics of the
sample stars, we are able to establish their inner/outer halo membership and
provide hints on their origin.  In addition, we can constrain the nature of the
Pop III progenitors of the lowest metallicity stars in our sample. This paper
is outlined as follows: the target selection and observations are described in
Section \ref{sect:obs}. Section  \ref{sec:three} discusses the determination of
stellar parameters. Our abundance analysis is presented in Section
\ref{sect:Method}. We discuss our results, chemical peculiarities, possible
progenitors, and kinematics in Section \ref{sec:discussions} . Our conclusions
are given in Section \ref{sec:conclusion}.

\section{Target Selection and Observations.}   
\label{sect:obs}

Our target stars were selected according to their corresponding Lick indices. We
used medium-resolution (R$ \sim$ 2,000) spectra from the third data release
(DR3\footnote{http://dr3.lamost.org}) of the Large Sky Area Multi-Object Fiber
Spectroscopic Telescope (LAMOST) survey \citep{2006ChJAA...6..265Z,
2012RAA....12..723Z, 2012RAA....12.1197C} to estimate their metallicities. 
High-resolution (R$ \sim$ 110,000) spectroscopic follow-up was
carried out, in order to confirm their low-metallicity nature and to explore
their elemental abundances in detail.

\subsection {High-resolution Observations}
\label{sect:obsred}

High-resolution spectroscopic observations for 12 stars were carried out using
the Levy Spectrometer \citep[optical echelle
spectrometer][]{2014PASP..126..359V} on the Automated Planet Finder (APF)
Telescope at Lick Observatory.  The observing setup included a 0.5" slit,
yielding a resolving power of R$ \sim$ 110,000 across wavelength range of
$3730$-$9989$ {\AA}\,, with an average S/N$ \sim$35 per pixel at 4500 {\AA}\,
(using 4$\times$30 minutes exposure times). For more information about the
target selection and observations we refer the reader to Paper I. These
high-resolution spectroscopic data were reduced through standard echelle data
reduction procedures (e.g., bias subtraction, flat-fielding), using the Image
Reduction and Analysis Facility \citep[IRAF;][]{1986SPIE..627..733T,
1993ASPC...52..173T}.  Six stars have been studied and reported in Paper I. In
this paper, we present the remaining stars. Table \ref{tab:obs} lists the
observational details of these stars. Our observed radial velocities (RV) were
measured by cross-correlating the high-resolution reduced spectra against
synthesized templates of the same spectral type, making use of the Mg I triplet
at 5160-5190 {\AA}\,. These values are also listed in Table \ref{tab:obs}.

\subsection{Equivalent Widths}
\label{sec:EWs_uncer}

We adopted a combined line list from \citet{2013AJ....145...13A} and
\citet{2013ApJ...769...57F}, for elements with Z $\leqslant 30$ (with the
exception of carbon and nitrogen, Z $=6$ and $7$, respectively). We rejected all
\ion{Fe}{1} lines with excitation potential lower than 1.2 eV, since these lines
may suffer from non-local thermodynamic equilibrium (NLTE) effects and yield
abundances that are not in agreement with the ones calculated from higher
excitation \ion{Fe}{1} lines \citep[e.g.,][]{1996A&A...305..245M,
2008ApJ...672..320C}. For the remaining elements considered in this study (e.g.,
\ion{Sr}{2}), we adopted atomic data from the linemake code
\footnote{https://github.com/vmplacco/linemake}.

The equivalent widths for the high-resolution spectra were measured using an
automatic IDL routine \citep{2015ascl.soft03003K} that fits Gaussian/Voigt
profiles to the selected lines, with expected errors of $\sim$ 1m{\AA}\,. We
visually examine each line and rejected blended lines or those with or
uncertain continuum placement. The equivalent widths values are listed in Table
\ref{tab:EWs}.

\begin{figure}[t!]
\epsscale{1.25}
\plotone{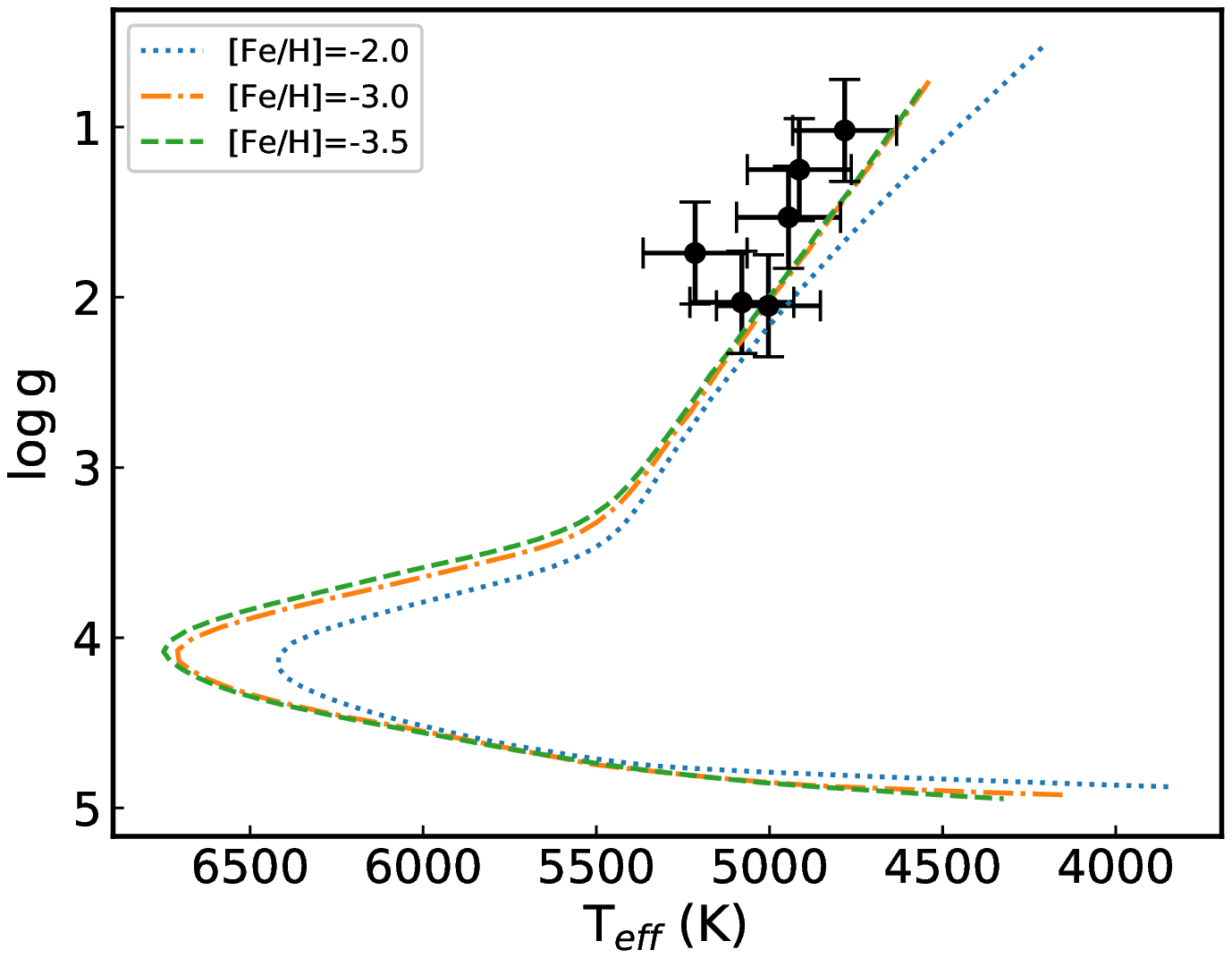}
\caption{The adopted \teff~and~\logg~ of our sample stars displayed with a
12~\Gyr~, \abund{\alpha}{Fe} = +0.4, and \abund{Fe}{H}= $-$2.5, $-$3.0, and
$-$3.5 Yale-Yonsei isochrones taken from \citet{2004ApJS..155..667D}.}
\label{fig:iso}
\end{figure}

\section{Stellar Parameters}
\label{sec:three}

The determination of atmospheric parameters, using photometric or spectroscopic
data, is essential to perform a detailed chemical abundances analysis
\citep{1999A&AS..140..261A, 2005ApJ...626..465R, 2010A&A...512A..54C,
2013ApJ...769...57F, 2017A&A...604A.129M}. However, it is well known that the
ionization equilibrium of \ion{Fe}{1} absorption features(spectroscopic method),
leads to underestimated \ensuremath{\log\,g} and \Teff\, \citep[e.g.,
see][and references therein]{2013ApJ...769...57F, 2017A&A...604A.129M}.

\textbf{We determined the atmospheric parameters for our target stars using a
number of available methods. We adopt the atmospheric parameters (corrected
values), derived based on the explicit method presented by
\citet{2013ApJ...769...57F}. This method corrects the usually expected
systematic offsets between the spectroscopic and photometric effective
temperatures. This exercise increases the determined spectroscopic \Teff\,
(initial values) up to several hundred degrees, and thus achieve better
agreement with values determined by other methods. These stellar atmospheric
parameters are listed in Table \ref{tab:stellar-param}.}

\subsection{Effective Temperature}

We determined the effective temperature for our sample stars by minimizing the
trend between the derived abundance and excitation potentials of \ion{Fe}{1}
lines. Moreover, we make use of the available colors (UCAC4,
\citealt{2013AJ....145...44Z} and 2MASS, \citealt{2006AJ....131.1163S}) along
with the empirical calibrations of \citet{2010A&A...512A..54C} to estimate
photometric \Teff. These \Teff\, values (photometric and spectroscopic) agree
within typical uncertainties ($\pm 150$ K). In addition, we cross-matched our
sample stars with Gaia DR2 \citep{2018A&A...616A...1G} and find that Gaia DR2
values agree well with our derived values.  These derived values together with
\Teff~taken from Gaia DR2 are listed in Table \ref{tab:stellar-param}. 

\subsection{Surface Gravity and Microturbulence}

We determined the surface gravities by minimizing the offset between the derived
abundances of \ion{Fe}{1} and \ion{Fe}{2} lines.  Moreover, we cross-matched our
sample stars against the distances catalogue of \citet{2018AJ....156...58B}, to
estimate \ensuremath{\log\,g} from the distance modulus.

\begin{eqnarray*}
\log \frac{g}{g_{\odot}} & = & \log \frac{M}{M_{\odot}}  + 4 \log \frac{T_{\rm eff}}{T_{\rm eff\odot}} \\
& & + 0.4(M_{bol} - M_{bol_{\odot}})\\
& & \\
M_{bol} & = & V+BC_{V}+5 \log \varpi+5
\end{eqnarray*}

\noindent where $M$ is the stellar mass, assumed to be $M/M_{\odot}= 0.8$,
$BC_{V}$ is the bolometric correction (determined based on
\citet{2014MNRAS.444..392C, 2018MNRAS.479L.102C,2018MNRAS.475.5023C}), $V$ is
the visual magnitude, $\varpi$ is the parallax, E(B-V) is the Color excess
(adopted from
\citet{1998ApJ...500..525S}\footnote{https://irsa.ipac.caltech.edu/applications/DUST}),
and $M_{bol}$ is the absolute bolometric magnitude.  The different
\ensuremath{\log\,g} values (distance modulus and spectroscopic) agree within
the typical uncertainty ($\pm$ 0.3\,dex).

Finally, we determined the microturbulence velocities ($\xi$) by removing any
trend between the calculated abundances of \ion{Fe}{1} lines and the associated
EWs. Figure \ref{fig:iso} shows the determined \ensuremath{\log\,g} as a
function of  \Teff, for our sample stars, overlaid with 12 \Gyr~Yale-Yonsei
isochrones \citep{2004ApJS..155..667D}, as a reference.  The error bars denote
$\pm 150$ K and $\pm 0.3$ cgs ($1~\sigma$ errors of \Teff~ and
\ensuremath{\log\,g}, respectively).

\begin{figure*}[t!]
\epsscale{1.25}
\plotone{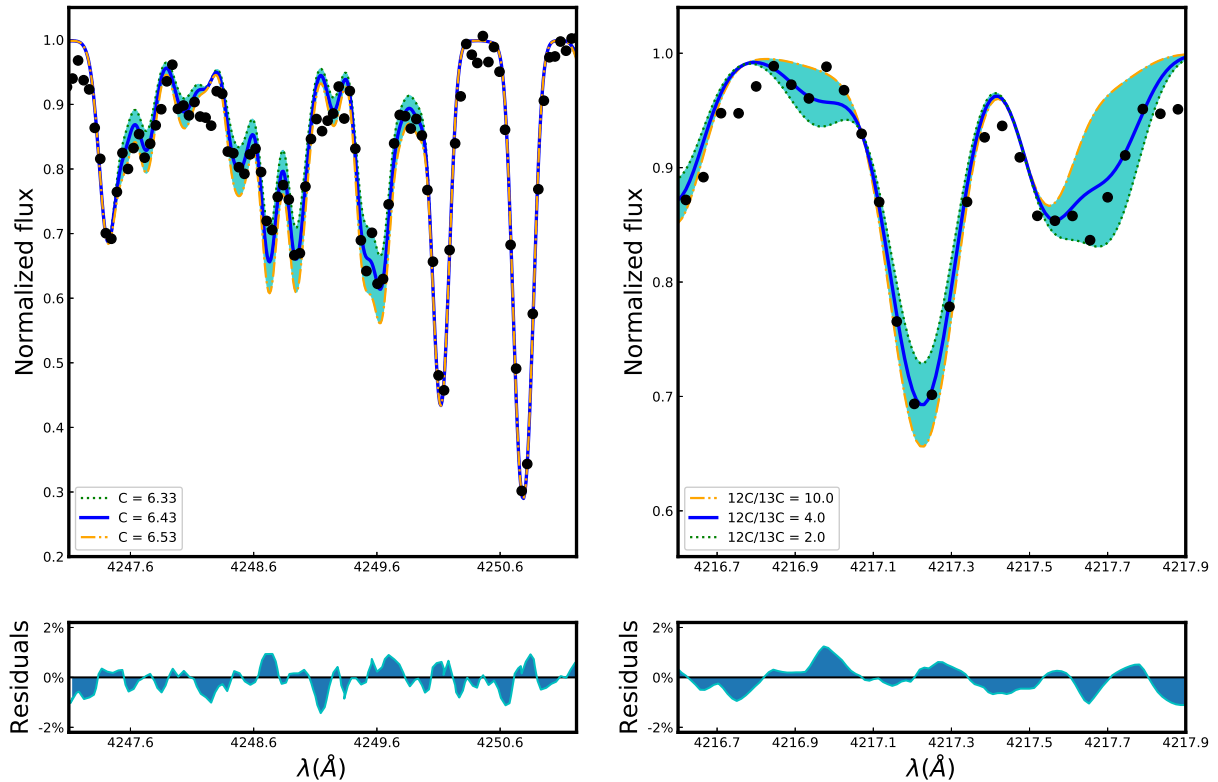}
\caption{Left panel: Section of the spectrum for J1630+0953 near the CH G band.
The filled black circles show the observed data, the solid blue line is the best
abundance fit, the green dotted line represents the upper abundance, and the
orange dot-dashed line represents the lower abundance. The cyan shaded area
encloses a 0.2 dex difference in log (C). The right panel shows the
determination of the carbon isotopic fractions (12C/13C). The filled black
circles show the observed data, the solid blue line is the best fractional fit.
The lower panels represent the residuals between the best fit and observed
data.}
\label{fig:light_Fe}
\end{figure*}

\begin{figure*}[t!]
\epsscale{1.2}
\plotone{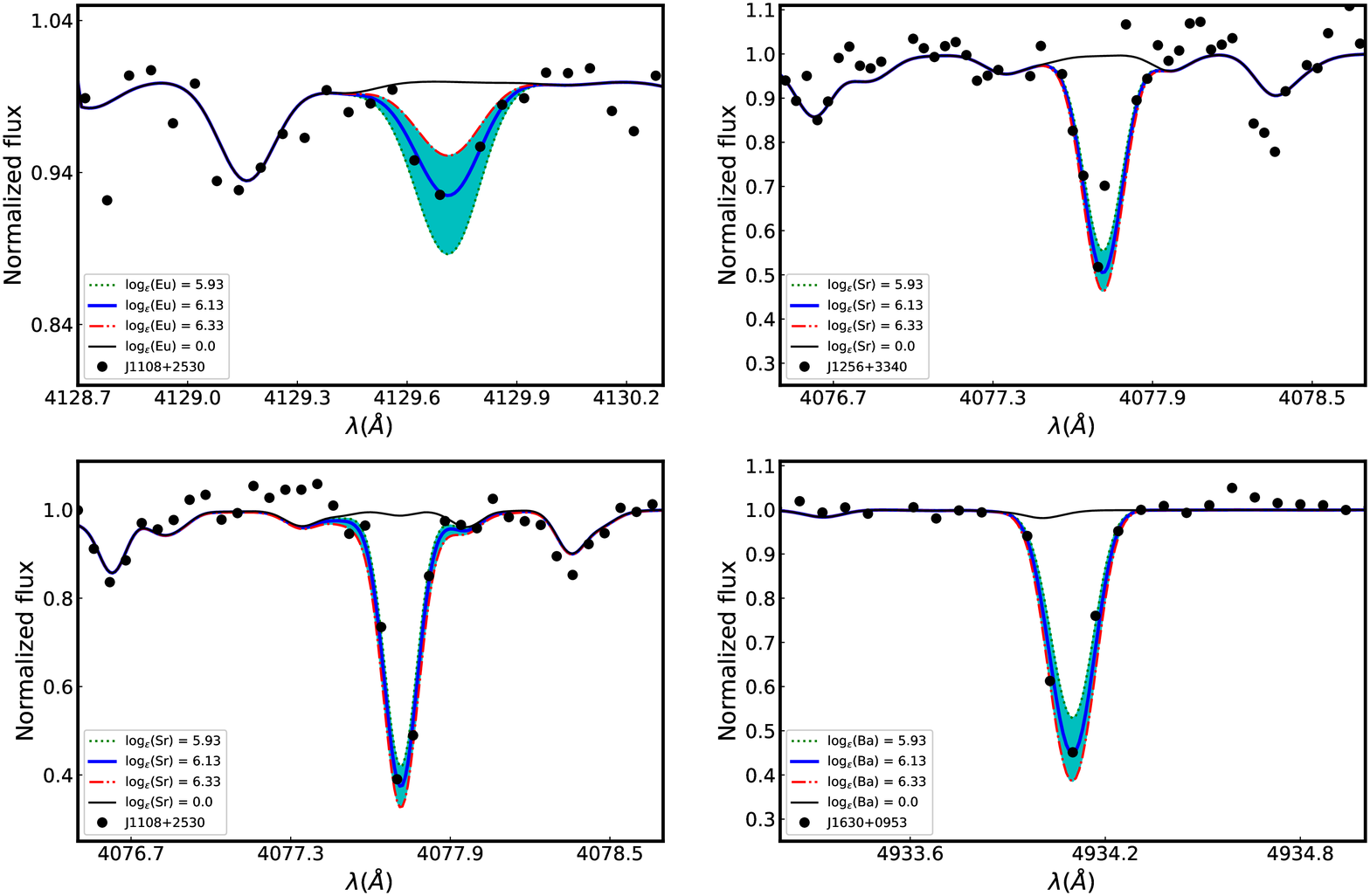}
\caption{Demonstration examples of our spectrum synthesis analysis. Black filled
circles denote portions of J1108+2530, J1256+3340, and J1630+0953 spectra. Blue
solid lines denote the best-fit, while the dotted and dash-dotted lines denote
the abundance variations used to estimate uncertainties.}
\label{fig:ncapture}
\end{figure*}

\section{Abundance Analysis}
\label{sect:Method}

We employed 1D stellar atmosphere models from \citet{2003IAUS..210P.A20C} and an
updated version of the stellar code MOOG
\citep{1973PhDT.......180S,2011AJ....141..175S, 2017AAS...23021607A} to
determine our LTE chemical abundances.
We used the measured EWs of the non-blended lines, with reliable continuum
normalization, to perform a standard LTE abundance analysis. At the same time,
we use spectral synthesis to determine abundances for blended lines, molecular
bands, and neutron-capture elements. Table \ref{tab:abund} presents the derived
LTE abundances. We adopted the solar abundances from \citet{2009ARAA..47..481A}
to calculate the final elemental abundances and [X/Fe] ratios.

\subsection{Carbon and Nitrogen}

We determined the carbon abundances in our sample stars from the molecular CH G
band around 4247 {\AA}\,.  Figure \ref{fig:light_Fe} shows the spectral
synthesis of the carbon abundance determination for J1630+0953 (upper left
panel). The black filled circles represent the observed data, the solid blue
line shows the best abundance fit, and the orange dashed and green dotted lines
show the upper and lower abundances fits, respectively.  We use these lines to
estimate the uncertainty of the carbon abundance determination. The lower panel
of Figure \ref{fig:light_Fe} shows residuals of $< 2\%$ between J1630+0953 data
and our best fit for the CH G region.

The surface chemical abundances of low mass stars ($\sim 1$ M$_{\odot}$) are
subject to change by the first dredge-up at the beginning of the red giant
branch (RGB). The effects of this stellar evolution phase can
be noticed (spectroscopically) by a lower observed ratio of $^{12}$C/$^{13}$C
and C/N \citep{1965ApJ...142.1447I}. The solar carbon isotopic ratio
\citep[$^{12}$C/$^{13}$C ratio $= 89,$][]{2009ARAA..47..481A} is expected to
decrease into values between 18 and 26 \citep[e.g.,][]{1994A&A...282..811C}, and
such enhancement in $^{13}$C can be used as an indicator of the degree of the
mixing processes in the outer layers of RGB stars.

To determine the carbon isotopic ($^{12}$C/$^{13}$C) ratios in our sample stars,
we fixed the carbon abundance (e.g., $\log \epsilon$ (C) = 6.13, for J1630+0953)
for the CH features around 4217 {\AA}\,, and then alter the isotopic ratio to
achieve a best fit ($^{12}$C/$^{13}$C = 4, for J1630+0953). Low
$^{12}$C/$^{13}$C ratios suggest that a significant amount of $^{12}$C has
been converted into $^{13}$C. In addition, a considerable amount of carbon
has been converted into nitrogen.  The upper right panel of Figure
\ref{fig:light_Fe} shows the determination of the carbon isotopic ratio. The
black filled circles represent the observed data, the solid blue line shows
the best carbon isotopic ratio fit, with lower (green dotted) and upper
(orange dashed) limits. The lower panel shows residuals of $< 2\%$ between
the observed data and the best fit ratio ($^{12}$C/$^{13}$C=4) at wavelength
$\sim 4217$ {\AA}\,.

We investigate the nitrogen bands (CN) at 4215 {\AA}\, and 6971{\AA} to
determine nitrogen abundances for our sample stars. Only the J1630+0953 spectrum
shows reliable features and consistent abundances. 

\subsection{Light Elements: from Na to Zn}

We used two lines (\ion{Na}{1} at 5889 and 5895 {\AA}\,) to determine Na
abundances, five lines (4703, 5172, 5183, 5528, and 5711 {\AA}\,) to derive Mg
abundances, and one reliable line at 4379.23 {\AA}\, to derive V abundances.
Moreover, we used 18 \ion{Ca}{1} lines, 26 Ti lines (\ion{Ti}{1} and
\ion{Ti}{2}), six \ion{Sc}{2} lines, five\ion{Cr}{1} lines, six \ion{Sc}{2}
lines, three \ion{Co}{1}lines, five \ion{Ni}{1} lines, and one \ion{Zn}{2} line
to determine the abundances for those species.

\subsection{Neutron-capture Elements}

We used spectral synthesis to derive abundances for the neutron-capture elements
in our sample stars: one \ion{Sr}{2} line at 4077 {\AA}\,, three \ion{Y}{2}
lines, four \ion{Ba}{2} lines, one \ion{Zr}{2} line, three \ion{La}{2} lines,
four \ion{Ce}{2} lines, two \ion{Pr}{2} lines, eight \ion{Nd}{2} lines, two
\ion{Sm}{2} lines, and one \ion{Eu}{2} line.  Figure \ref{fig:ncapture}
demonstrates our spectral-synthesis fittings. The uncertainties are estimated by
using same method described for carbon.

\begin{figure}[t!]
\epsscale{1.2}
\plotone{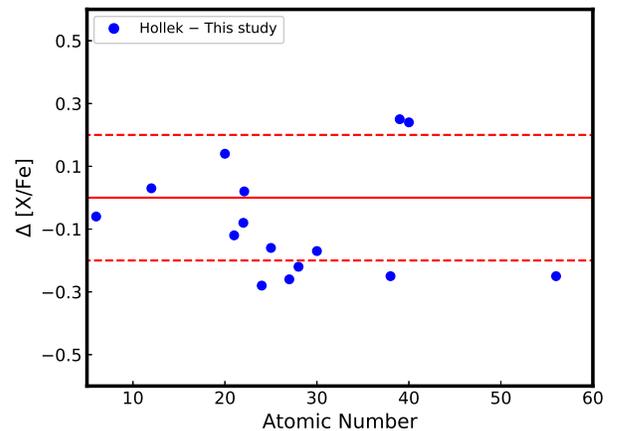} 
\caption{Elemental abundance differences of J0326+0202, as a function of the
atomic number. The solid line refers to zero elemental abundance difference and
the dashed lines refer to $\pm 0.2$ dex deviations.}
\label{fig:J0326}
\end{figure}

\subsection{Uncertainties of Stellar Parameters and Abundances}
\label{subsec:abuns_uncer} 

We tested the robustness of the determined elemental abundances by altering one
atmospheric parameter (\teff, \logg, and $\xi$) at a time within their associated
uncertainty (100 K, 0.3 dex, and 0.3 km s$^{-1}$, respectively).  Table
\ref{tab:errors} shows the effect of this procedure on the abundances
of J1630+0953, as an illustrative example.  The derived chemical abundances of
J1630+0953 have an overall average uncertainty of 0.2 dex. Nonetheless, we
neglect the uncertainties arising from the choice of the atmospheric model
\citep[see][]{2003IAUS..210P.A20C} and the EWs measurements, because they result
in negligible systematic errors, compared with the errors on the atmospheric parameters.
\citep[e.g.,][]{2005Natur.434..871F, 2006ApJ...639..897A}.

The elemental abundances of J0326+0202 have been discussed by
\citet{2011ApJ...742...54H}, who analyzed a high-resolution spectrum (R $\sim$
35,000) obtained using the Magellan/MIKE. Figure \ref{fig:J0326} shows the
differences between the abundances in common derived in this work and
\citet{2011ApJ...742...54H}. We attribute the differences to the use of
different stellar parameters \citep[][\textbf{used \teff=4775 K, \logg=1.20 dex,
\metal=$-$3.32 dex, and $\xi$= 1.80 kms$^{-1}$}]{2011ApJ...742...54H}, different
atomic data, and different techniques (equivalent width vs. spectrum synthesis).

\section{Results and Discussion}\label{sec:discussions}
  
\subsection{Chemical Abundance Comparison with Literature Data}
 
\begin{figure*}[t!]
\epsscale{1.2}
\plotone{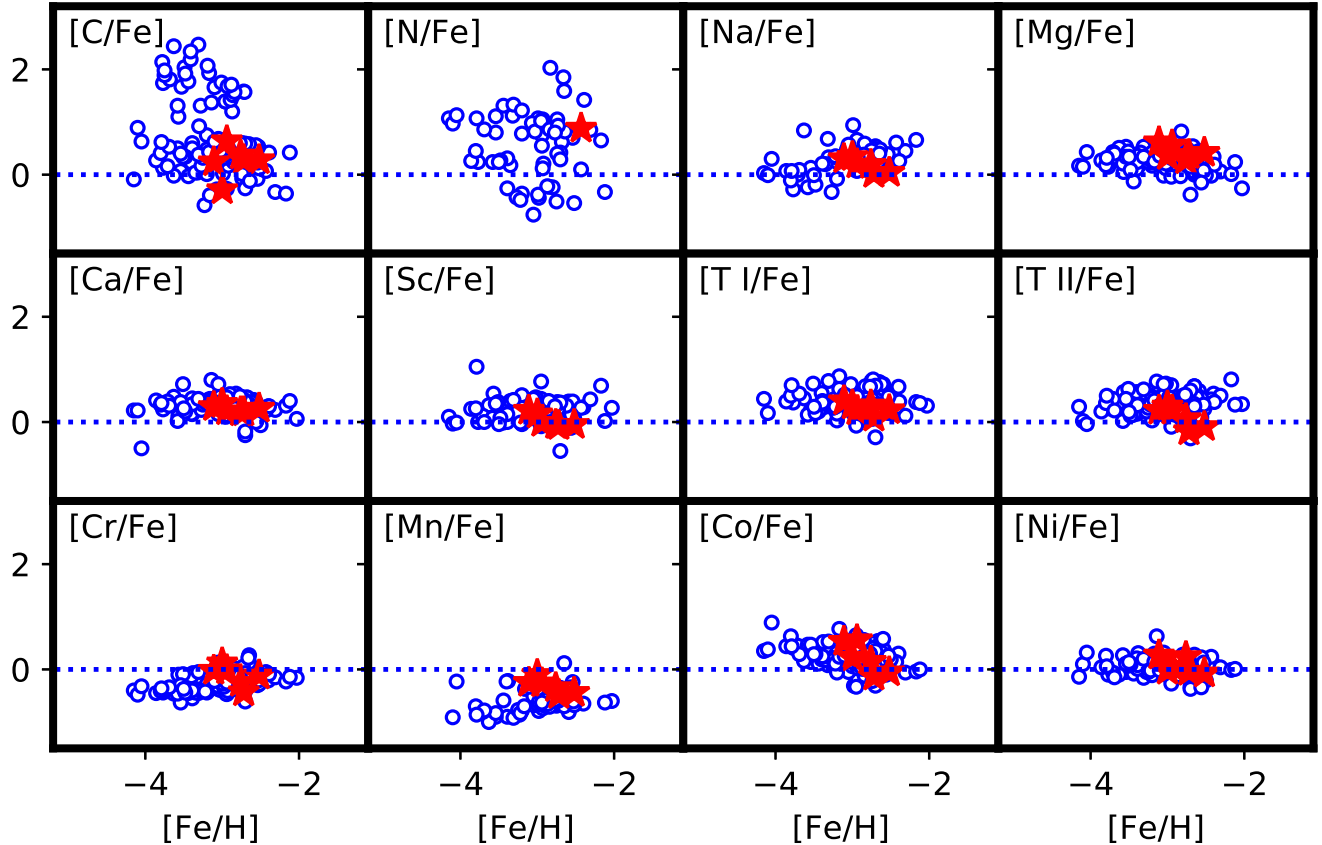} 
\caption{Selected light-elements abundances of our sample stars (red filled
stars) overlaid with \abund{X}{Fe} of carbon-normal metal-poor stars adopted
from \citet{2013ApJ...762...26Y} (blue open circles). Our derived \abund{X}{Fe}
show no significant differences compared to the literature data. \abund{C}{Fe}
values do not represent the carbon abundances of the natal gas.}
\label{fig:abund}
\end{figure*}

%\begin{figure*}[t!]
%\epsscale{1.2}
%\plotone{Francois.eps} 
%\caption{Neutron-capture elements abundance ratios, as a function of metallicity. Red filled stars denote our sample stars, blue open circles denote data adopted from \citet{2007A&A...476..935F}. No significant differences compared with \citet{2007A&A...%476..935F} are found.}
%\label{fig:dtrans}
%\end{figure*}

To identify any possible chemical peculiarities in our sample, we compared our
determinations with data from the literature. Figure \ref{fig:abund} shows some
of the light-element abundances, compared to C-normal stars taken from
\citet{2013ApJ...762...26Y}. The detectable light-elements have similar
abundance ratios to those reported in literature halo stars, suggesting that our
sample stars have been likely formed from a well-mixed gas cloud.

All our sample stars have been evolved into the red giant branch (see Figure
\ref{fig:iso}), suggesting that some internal mixing occurred and thus the
atmospheric carbon abundances have been altered.  \citet{2014ApJ...797...21P}
studied a sample of 505 metal-poor stars and developed a correction procedure that
recovers the initial atmospheric carbon abundance. These corrections are
listed in Table \ref{tab:abund}.  We obtained high $\abund{C}{Fe}_{corr}$ values
for J1630+0953 and J2216+0246 ($\abund{C}{Fe}_{corr}$ = +0.62 and 0.42 dex,
respectively), which brings the final abundances to \abund{C}{Fe}=1.26 and
0.70. From these final values and the definition of the CEMP
stars adopted from \citet[][CEMP, \abund{C}{Fe} $\geq
0.7$]{2007ApJ...655..492A}, we classify J1630+0953 and J2216+0246 as
carbon-enhanced metal-poor stars, while the rest of our program stars are
carbon-intermediate.

The substantial depletion of carbon in J1630+0953 and J2216+0246 suggests that
the chemical patterns observed in these two stars should be accompanied with
nitrogen enhancement. However, we were only able to detect nitrogen in the
spectrum of J1630+0953 (\abund{N}{Fe}=0.88), which is in line with its
evolutionary status, while J2216+0246 spectrum show no reliable CN features.  

\begin{figure}[t!]
\epsscale{1.2}
\plotone{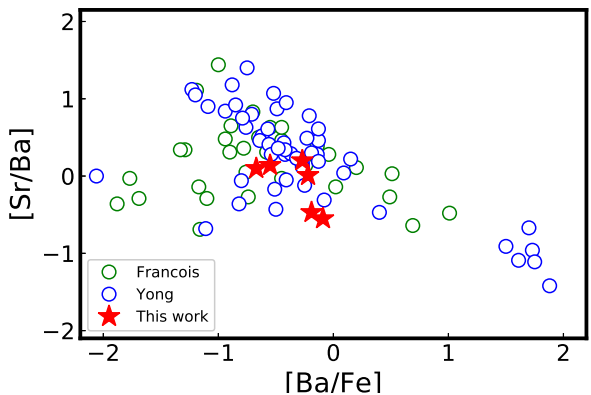} 
\caption{The determined abundance ratios \abund{Sr}{Ba} of our sample stars (red
filled stars), versus \abund{Ba}{Fe}. The green and blue open circles denote
literature data adapted from \citet{2007A&A...476..935F} and
\citet{2013ApJ...762...26Y}, respectively.}
\label{fig:Sr_Ba}
\end{figure}

We have determined chemical abundances for up to nine neutron-capture elements
in our sample stars. Of particular interest, the observed Sr and Ba abundances
could help us better understand their nucleosynthesis pathway(s), considering
that Sr (first s-process peak) may be synthesized by the main or weak s-process,
while Ba (second s-process peak) primarily synthesized by the main s-process
\citep{2003ApJ...588.1099Q, 2008ApJ...687..272Q, 2011A&A...530A.105A,
2014ApJ...797..123H}. Furthermore, \citet{2009ApJ...696..797C,
2011ApJS..197...17C} and \citet{2012ApJ...747....2L} predicted that low and
high \abund{Sr}{Ba} ratios could help discriminate between low-mass and
massive metal-poor AGB stars, respectively. This further suggests that the
production of Sr and Ba are linked to various astrophysical sites. On the
other hand, \citet{2013ApJ...766L..13A} suggested that these elements could
be synthesized in the same event, and the observed \abund{Sr}{Ba} ratios can
be explained by the stars collapse time into a black-hole. 

Figure \ref{fig:Sr_Ba} shows the \abund{Sr}{Ba} abundance ratios of our program
stars (filled red stars) and carbon-normal stars from
\citet{2007A&A...476..935F} (green open circles) \citet{2013ApJ...762...26Y}
(blue open circles), as a function of \abund{Ba}{Fe}. All of the sample stars
show no significant differences from the trends presented by
\citet{2007A&A...476..935F} and \citet{2013ApJ...762...26Y}. However, abundances
from elements in the first s-process peak in J0326+0202 (\abund{Sr}{Ba} = 0.14)
and J1413+1727 (\abund{Sr}{Ba} = 0.20) appear to be enriched more than the
second peak, thus an additional process is required to explain these high
\abund{Sr}{Ba} ratios.  J1630+0953 and J2216+0246 satisfied the CEMP definition
\citep[\abund{C}{Fe} $\geq 0.7$;][]{2007ApJ...655..492A}. In addition,  we
determined low Ba abundances (\abund{Ba}{Fe}$<$ 0.0) for these stars (see Table
\ref{tab:abund}).  Therefore, J1630+0953 and J2216+0246 can be classified as
CEMP-no stars \citep[see][Section \ref{sec:intro}, and Figure
\ref{fig:Sr_Ba}]{2005ARA&A..43..531B, 2007ApJ...655..492A, 2014ApJ...797...21P}. 

\begin{figure*}[t!]
\epsscale{1.2}
\plotone{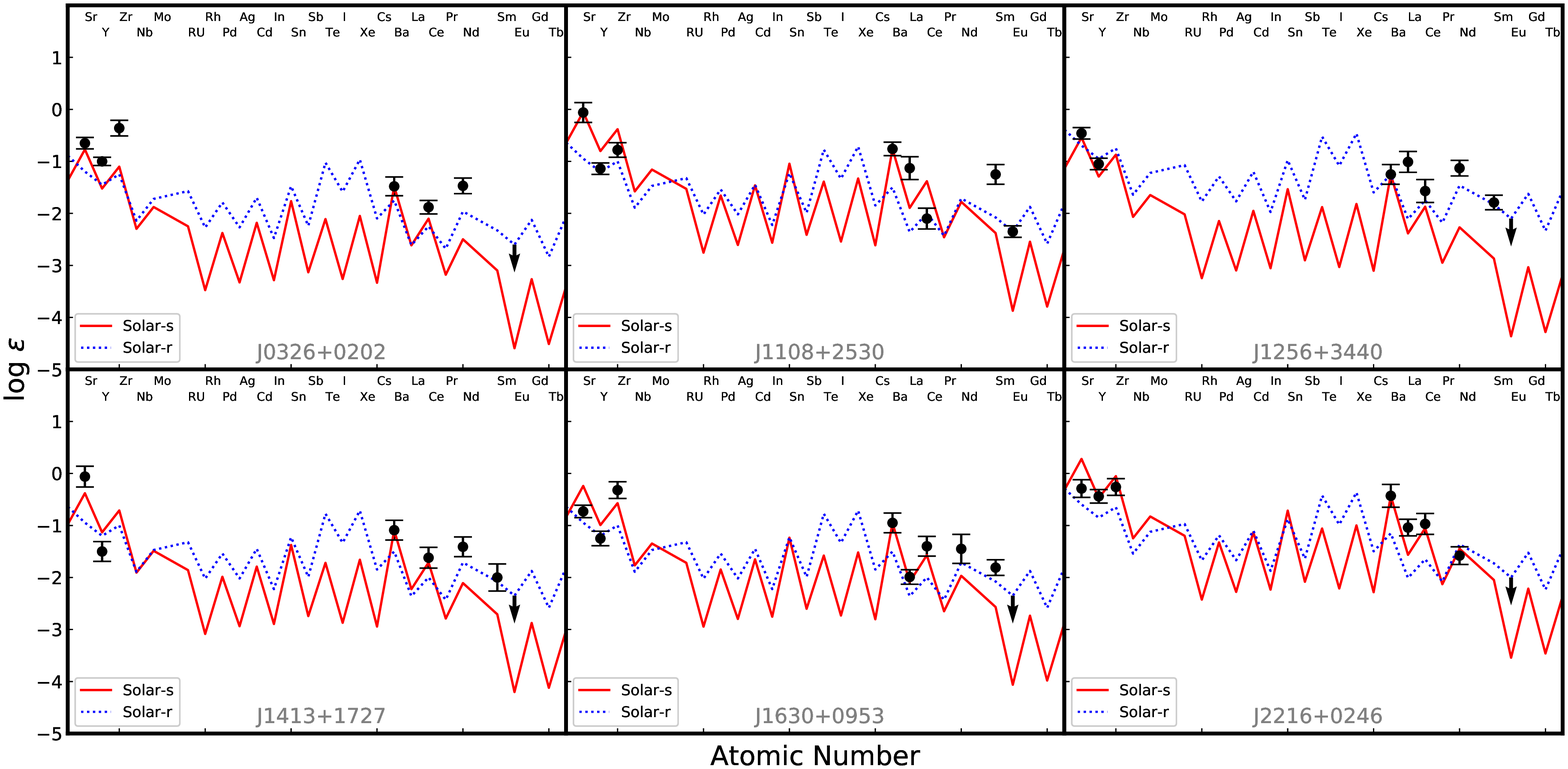} 
\caption{Elemental abundance patterns for the heavy element distribution in our
sample stars (filled circles denote detections, and arrow denotes $3 \sigma$
upper limits derived from \ion{Eu}{2} line), compared with scaled solar system
r- and s-process components adopted from \citet{2000ApJ...544..302B}. The solar
system r-process patterns are scaled to match Eu, and the solar system s-process
patterns are scaled to match Ba.}
\label{fig:solar_r_s-process}
\end{figure*}

\begin{figure*}[t!]
\epsscale{1.2}
\plotone{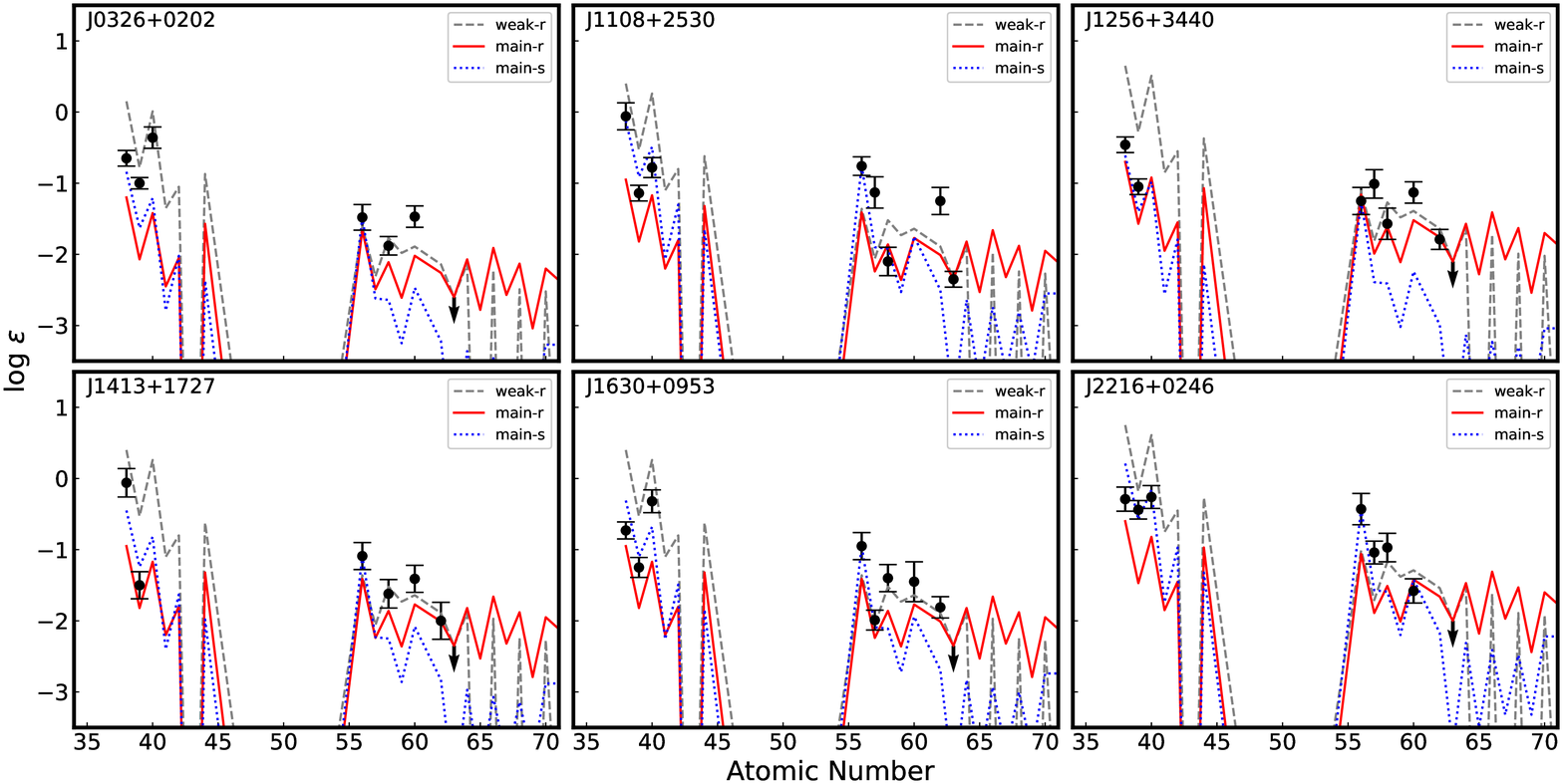} 
\caption{Neutron-capture element patterns in our sample stars. Filled circles
denote detections, and arrow denotes $3 \sigma$ upper limits derived from
\ion{Eu}{2} line. The dashed gray line denotes the observed abundances for the
neutron-capture elements in HD122563 \citep{2006ApJ...643.1180H,
2012ApJS..203...27R}. The solid red line denotes the observed abundances for the
neutron-capture elements in CS 22892–052 \citep{2003ApJ...591..936S,
2009ApJS..182...80S, 2009ApJ...698.1963R}. The dotted blue line denotes
predicted yields from s-process nucleosynthesis in TP-AGB stars
\citep{2008ARA&A..46..241S, 2011MNRAS.418..284B}. The physical meaning of these
three lines are discussed in the text.}
\label{fig:weak_main}
\end{figure*}

By comparing the observed neutron-capture abundance patterns in our sample
stars with the solar system (SS) r- and s-fractions, we can gain insights into
the nature of their heavy element enrichment. Figure \ref{fig:solar_r_s-process} shows the
heavy element abundance patterns of our sample stars compared with the SS r- and
s-fractions, adopted from \citet{2000ApJ...544..302B}. The SS s-process
components are normalized to match the observed Ba (solid line), and the SS
r-process patterns are normalized to match the observed Eu (dotted line). The
derived abundances for Sr, Y, and Zr (within observational errors) in J0326+0202
seem to be inconsistent with the SS distributions. These discrepancies require
an additional process(es) to interpret the overall neutron-capture abundance
pattern observed in this star. However, these elements seem to be consistent
with the SS s-process distributions for the rest of our sample stars. Heavier
elements (Z $\geqslant 56$) in J0326+0202, J1256+3440,  J1413+1727, and
J1630+0953 are in better agreement with the normalized SS r-process than the
normalized SS s-process. The heavier elements in J2216+0246 seem to favor the
normalized SS s-process. The abundance pattern in J1108+2530 is not in agreement
with the SS abundance pattern.

Figure \ref{fig:weak_main} shows the heavy element abundance patterns in our
sample stars compared with abundance patterns of HD 122563 \citep[the dashed
gray line,][]{2006ApJ...643.1180H, 2012ApJS..203...27R}, CS 22892–052 \citep[the
solid red line,][]{2003ApJ...591..936S, 2009ApJS..182...80S,
2009ApJ...698.1963R}, and predicted yields from s-process nucleosynthesis in
TP-AGB stars \citep[the dotted blue line,][]{2008ARA&A..46..241S,
2011MNRAS.418..284B}. \textbf{These patterns can be used as weak component of
the r-process (HD 122563), main component of the r-process (CS 22892–052), and
main component of the s-process (TP-AGB yields) representatives
\citep{2014ApJ...784..158R}}. The abundance pattern of HD122563 and CS 22892–052
are renormalized to match the observed Eu, and the dotted blue line is
renormalized to match the observed Ba. The abundance patterns for the heavy
elements (Z $\geqslant 56$) observed in the sample stars are in better agreement
with the weak component of the r-process (represented by HD 122563), with a
partial s-process contribution in J1108+2530, J1630+0953, and J2216+0246.

\begin{figure}[t!]
\epsscale{1.2}
\plotone{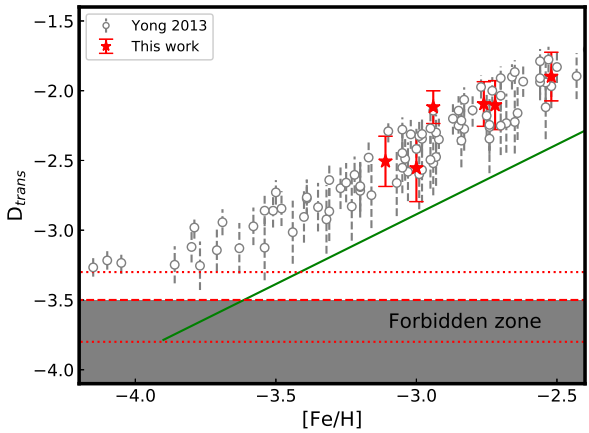} 
\caption{Dtrans as a function of \abund{Fe}{H} for our target stars, compared
with \citet{2013ApJ...762...26Y}. The solid green line denotes the scaled solar
pattern, the red dashed line denotes the limit of Dtrans based on the model
described in Frebel et al. (2007), and the red dotted lines show the uncertainty
of the model. The shaded area is the Forbidden Zone, where there is insufficient
C and O induced cooling for low-mass star formation.}
\label{fig:dtrans}
\end{figure}
  
In addition to the possible nucleosynthesis pathway(s) suggested above, it is
now commonly acknowledged that the formation of Pop II low-mass stars requires
additional coolants compared to the formation of Pop III massive stars
\citep[e.g.,][]{2006ApJ...652....6Y, 2013ASSL..396..103G, 2015ComAC...2....3G}.
At present, a widely acceptable scenario tries to explain the transition from
Pop III to Pop II stars by the notion of the fine-structure cooling of the
\ion{C}{2} and \ion{O}{1} lines \citep[e.g.,][]{2003Natur.425..812B,
2006ApJ...643...26S}. Therefore, C- and O-poor stars can place crucial
constraints on the conditions under which metallicity of the ISM the Pop II
stars may have been formed. 

\citet{2007MNRAS.380L..40F} suggested a criterion for the transition to Pop II
(``transition discriminant" $-$ D$_{trans} = -3.5$) , which combines logarithmic
abundance ratios of \abund{C}{H}  and \abund{O}{H}. Figure \ref{fig:dtrans}
shows the calculated D$_{trans}$ for our sample stars and carbon-normal stars
adopted from \citet{2013ApJ...762...26Y}. The green solid line denotes the solar
D$_{trans}$ values adopted from \citet{2009ARAA..47..481A}, the red dashed and
the red dotted lines denote the limit of D$_{trans}$ and its associated
uncertainty, respectively, and the gray shaded area represents the Forbidden
Zone. Even though the empirical formula of the D$_{trans}$ requires both
abundances (\abund{C}{H} and \abund{O}{H}), it is possible to calculate
D$_{trans}$ where only  one of these abundance ratios is known. Here we assume
that \abund{O}{Fe} $= 0.88 \pm 0.28$, following the linear relation between the
\abund{C}{O} and \abund{C}{Fe} suggested by \citet{2004A&A...416.1117C}. We can
notice that our sample stars are located above the solar D$_{trans}$ values and
the Forbidden Zone, suggesting that they may have likely formed from a gas cloud
exhibiting fine-structure cooling process.

\begin{figure*}[t!]
\epsscale{1.2}
\plotone{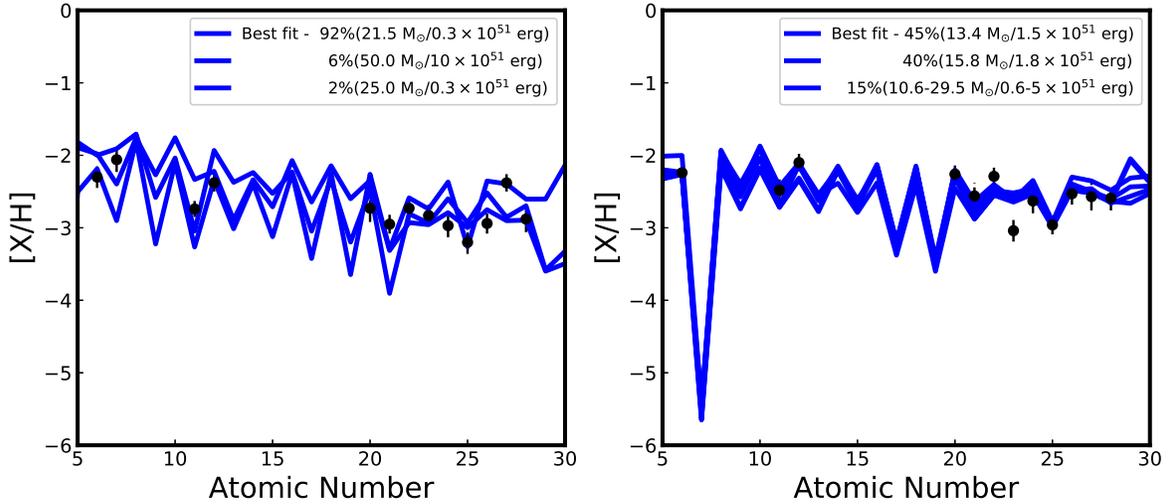} 
\caption{The determined \abund{X}{H} abundance ratios of J1630+0953 (left panel)
and J2216+0246 (right panel) as a function of atomic number, overlaid with
simulated abundance patterns by several best fit models, transparent by their
fractional appearance. The best fits and their properties are discussed in the
text.}
\label{fig:starfit}
\end{figure*}

\subsection{J1630+0953 and J2216+0246 Possible Progenitors}

Under the assumption that J1630+0953 and J2216+0246 are CEMP-no stars, we
decided to gain insights into their putative progenitors, by comparing their
abundance patterns with theoretical predictions from
\citet[][]{2010ApJ...724..341H}.  For the convenience of the reader, it is worth
mentioning that this grid does not consider rotation and has a $\chi^{2}$
matching algorithm over 16,800 models.  The parameter space includes progenitor
masses (10-100 M$_{\odot}$), explosion energies ($0.3$-$10 \times 10^{51}$ erg),
and mixing factor ($f_{mix}$) ranging from no mixing to nearly complete mixing
\citep[see][for more information]{2010ApJ...724..341H}. 

\begin{figure}[t!]
\epsscale{1.2}
\plotone{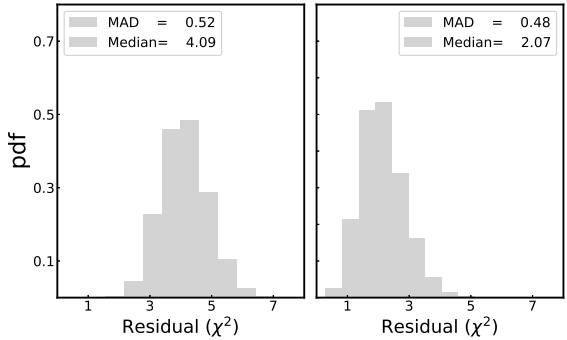} 
\caption{Posterior Distributions for $\chi^{2}$, of the 10,000 simulations, for
J1630+0953 (left panel) and J2216+0246 (right panel). The median and median
absolute deviation (MAD) are shown in legends.}
\label{fig:Starfit_histo}
\end{figure}

We sampled $10^{4}$ sets of the determined chemical abundances of J1630+0953 and
J2216+0246, assuming a normal distribution. To facilitate this exercise, we use the
determined log\,$\epsilon$(X) of J1630+0953 and J2216+0246 as the central values
and dispersions given by the associated uncertainties (see references to Table
\ref{tab:abund}). This allowed us to generate $10^{4}$  abundance patterns
for each star. We use the publicly available \texttt{STARFIT} code
\citep{2010ApJ...724..341H} to find the progenitor mass and explosion energy for
the $10^{4}$ abundance patterns. Figure \ref{fig:starfit}
shows the determined abundances of J1630+0953 and J2216+0246 (black filled
circles), with error bars representing their associated uncertainties,
overlaid with abundance patterns generated by several best fit models.
Figure  \ref{fig:Starfit_histo} shows posterior distributions for the mean
squared residuals of the $10^{4}$ fittings, for both stars. Legends show the
median value and the median absolute deviation (MAD). The MAD can be used as a
robust estimator on how the data spreads out. In other words, the larger the
MAD, the greater the variability in $\chi^{2}$.

For J1630+0953, an SN model with mass 21.5 M$_{\odot}$ and explosion energy $0.3
\times 10^{51}$ erg was the most frequent model (92 $\%$) to fit the generated
abundance patterns. Another model with mass 50.0 M$_{\odot}$ and explosion
energy $10.0  \times 10^{51}$ erg fit 6 $\%$ of the generated abundance
patterns, while the remaining of the $10^{4}$ generated abundance patterns (249)
had best fit models with mass 25.0 M$_{\odot}$ and explosion energy $0.3 \times
10^{51}$ erg. 

For J2216+0246, a larger number of SN models were able to fit our $10^{4}$
generated abundance patterns. About 45 $\%$ (4532) of the generated abundance
patterns had best fit with an SN model with mass of 13.4 M$_{\odot}$ and
explosion energy of $1.5 \times 10^{51}$ erg. Another SN model with mass 15.8
M$_{\odot}$ and explosion energy $1.8 \times 10^{51}$ erg was the best fit of 40
$\%$ (3976) of the generated abundance patterns; while 21 different SN models
correspond to the best fit for the rest of the $10^{4}$ generated abundance
patterns (1492). In total, 23 different SN models were the best fits for the
$10^{4}$ generated abundance patterns of J2216+0246, in the mass range 10.6-29.5
M$_{\odot}$ and explosion energies $0.6$-$5 \times 10^{51}$ erg. The abundance
patterns simulated by these 23 models are shown in the right panel of Figure
\ref{fig:starfit}, and color-coded by their fractional appearance.

Regarding the most frequent SN model in the exercise of J1630+0953, it can be
seen through either visual checking or $\chi^{2}$, that the carbon (Z=6) and
nitrogen (Z=7) abundances are well reproduced (within$\sim 2~\sigma$). Sodium
(Z=11) is overproduced, while magnesium (Z=12) is well reproduced (within$\sim
1~\sigma$). Elements from calcium to iron (Z=20-26) are also well reproduced
(within $\sim 2~\sigma$). The cobalt (Z=27) abundance is underproduced, which is
not unusual for theoretical models, where an SN model with higher mass,
explosion energy, and mixing factor may result better Co fitting, although this
could worsen other elements fittings \citep[see][and references
therein]{2014ApJ...785...98T}, while the Nickel (Z=28) is well reproduced. The
predicted yields in  $\sim~85 \%$ of the SN models reproduced the observed
abundances for J2216+0246  within$\sim 2~\sigma$. 

In general, J1630+0953 has possible progenitor with 21-25 M$_{\odot}$ stellar
mass and explosion energy $0.3 \times 10^{51}$ erg, while the mass and
explosion energy of the possible progenitors of J2216+0246 are somewhat lower
(10.6-29.5 M$_{\odot}$, $0.6$-$5 \times 10^{51}$ erg). Recently,
\citet{2018ApJ...857...46I} compared the abundance patterns of 219 EMP stars
with supernova yields of metal-free stars to find that the best fitting
progenitor SNe of most EMP stars are PopIII stars in the range 15-25
M$_{\odot}$. \textbf{The analysis presented in this paper supports this
hypothesis and suggests that the peak around 20 M$_{\odot}$ may reflect the Pop
III initial mass function and more massive SN might be more energetic that their
ejecta escape from the halo and would never been incorporated into the next
generation of stars.}

\subsection{Kinematics and Dynamics} \label{sec:kinematics}

The full space motion of our sample stars can be derived by combining positions
($\alpha$, $\delta$), proper motions ($\mu_{\alpha}\cos\delta$, $\mu_{\delta}$),
available in Gaia DR2 \citep{2018A&A...616A...1G}, the line-of-sight velocities
($V_r$), derived from our high-resolution spectra (see Section
\ref{sect:obsred}), and a Galactic potential model. Errors are provided in Gaia
DR2, thus the inversion of the parallax ($\varpi$) to calculate the stellar
distance is not appropriate \citep[see][for a recent
discussion]{2018A&A...616A...9L}.  Therefore, we adopted distances from
\citet[][]{2018AJ....156...58B}, who inferred distances to all stars, with
published parallaxes, in Gaia DR2 using a weak distance prior. Gaia DR2 ID
source, positions, proper motions, distances, and the associated uncertainties
for these quantities, are listed in Table \ref{tab:kinematics_input}. 

\begin{figure*}[t!]
\epsscale{1.2}
\plotone{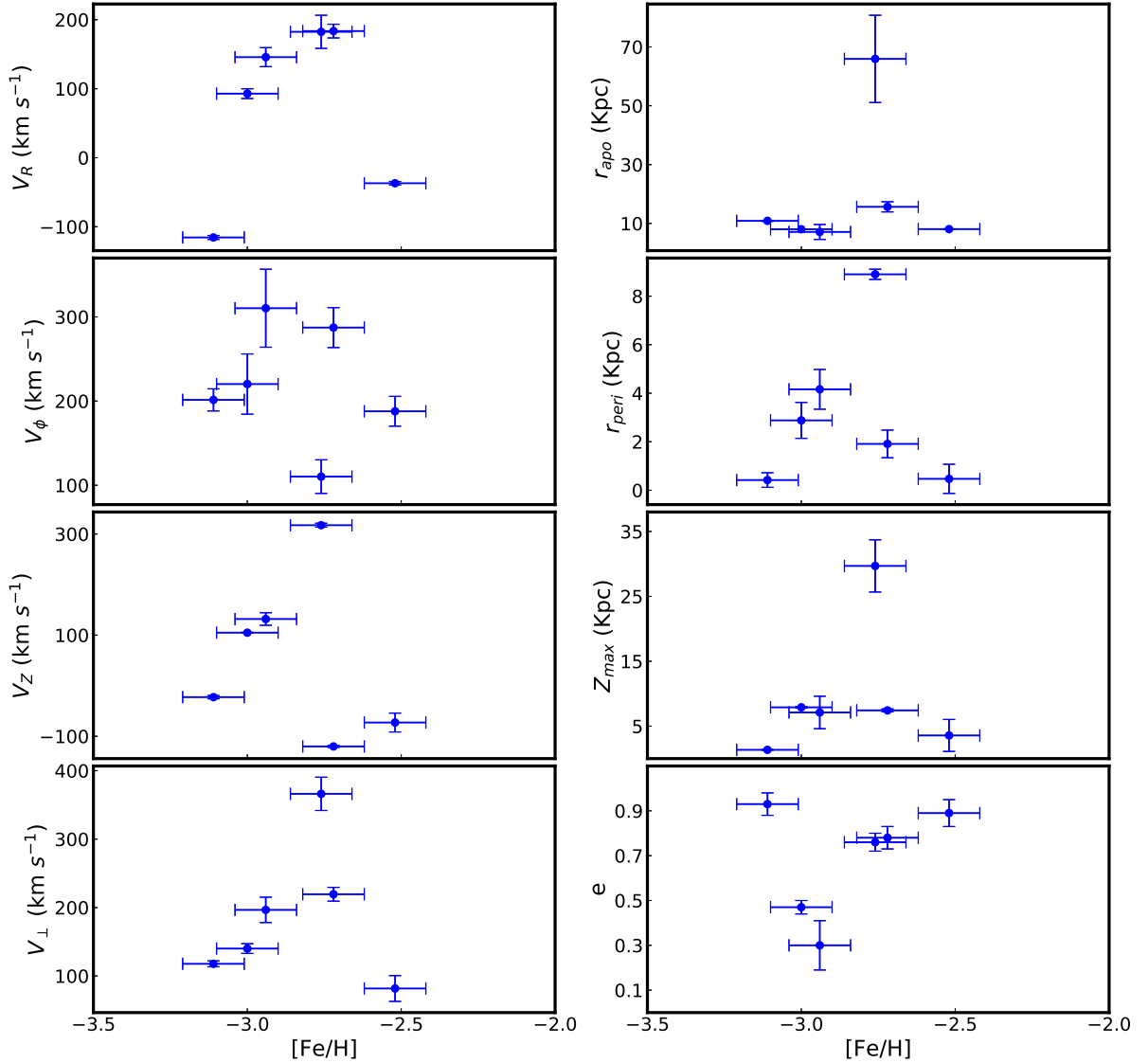} 
\caption{The left panel shows the Galactic velocities and the right panel shows
the orbital parameters for our sample stars as functions of \abund{Fe}{H}. The
y-axis error bars denote the 16$^{\rm th}$ and 84$^{\rm th}$ percentiles, while
the x-axis error bars represent a typical \abund{Fe}{H} uncertainty (0.10
$\sim$dex).}
\label{fig:Kinematics}
\end{figure*}

We sampled $10^{4}$ sets of the observed astrometric quantities (RA, DEC,
$\varpi$, RV, $\mu_{\alpha}\cos\delta$, $\mu_{\delta}$) from the measurement
errors of each quantity for each star in our sample (see references to Table
\ref{tab:kinematics_input}). We assume that the Sun has an offset above the
Galactic midplane of $z_{\odot}= 25$ pc \citep{2008ApJ...673..864J}, $R_{\odot}
=$~8.2~kpc \citep{2016ARA&A..54..529B} as the distance from the Galactic center,
circular velocity $v_0 =$~232.8~\mbox{km~s$^{\rm -1}$} at the Solar position
\citep{ 2017MNRAS.465...76M}, and solar peculiar motion $(U_\odot, V_\odot,
W_\odot) =$ (11.1, 12.24, 7.25)~\mbox{km~s$^{\rm -1}$}
\citep{2016ARA&A..54..529B}.  We calculate the Galactocentric Cartesian
($X_{GC}, Y_{GC}, Z_{GC}$) coordinates as follows:

\begin{eqnarray*} \label{eq:xyz}
X_{GC}  & = & R_{\odot} - d \cos (b) \cos (\ell) \\
Y_{GC}  & = &- d \cos (b) \sin (\ell)  \\
Z_{GC}  & = & d \sin (b) + z_{\odot}\\
\end{eqnarray*}

We calculated and corrected the Galactic space-velocity components $(U, V, W)$
using the Astropy Galactocentric frame package \citep{2013A&A...558A..33A,
2018AJ....156..123A}: U (positive toward the Galactic center), V (positive in
the direction of Galactic rotation) and W (positive toward the North Galactic
Pole).  Moreover, we define the angle $\phi = \tan^{-1}(Y_{GC}/X_{GC})$ and then
calculate the cylindrical velocities components for our sample as follows:

 \begin{eqnarray*} \label{eq:xyz}
V_{R}    & = & U \cos (\phi) + V \sin (\phi)\\
V_{\phi} & = & U \sin (\phi) - V \cos (\phi) \\
V_{z}     & = & W \\
\end{eqnarray*}

We adopted the \texttt{MWPotential2014} as a Galactic potential model
\citep[see][for more information]{2015ApJS..216...29B} to integrate the
corresponding stellar orbits, apocentric ($\rapo$) and pericentric ($\rperi$)
radii, the maximum offset from the Galactic midplane ($\zmax$), and
eccentricity, defined as $e = (\rapo - \rperi) / (\rapo + \rperi)$. In addition,
we derived the total orbital energy, defined as $E = (1/2) \vector{v}^2 +
\Phi(\vector{x})$ and the angular momentum in the vertical direction, defined as
$L_z = R \times V_{\phi}$, where $R$ denotes the distance from the Galactic
center projected onto the disk plane.

Table \ref{tab:UVW} lists, for each star in our sample, the calculated median
for the Galactic positions $(X_{GC},~Y_{GC},~Z_{GC})$, Galactic velocities
$(U,~V,~W,~V_{R},$ and $~V_{\phi})$, as well as $V_{\perp}$, defined as
$(V_{R}^{2} + V_{z}^{2})^{1/2}$. Table \ref{tab:energy} lists the calculated
median for the orbital parameters, energy and angular momentum. We also
provide an estimated $\sim 1~\sigma$ uncertainty of each quantity in our
calculations, defined as the differences between the median and 84$^{\rm
th}$ percentiles (superscript), and the differences between the median and
16$^{\rm th}$ percentiles (subscript).

Figure \ref{fig:Kinematics} shows the behavior of the calculated velocities and
orbital properties, for our sample stars, as a function of \abund{Fe}{H}. Error
bars denote typical \abund{Fe}{H} uncertainty (x-axis) and the 16th and 84th
percentiles (y-axis). It is possible to see that 66 $\%$ of our sample is moving
away (V$_{R} >0$) from the Galactic center, 100$\%$ on prograde (V$_{\phi} >0$)
orbits, and 50 $\%$ moving north (V$_{Z} >0$), as they pass through the Galactic
disk. The right panel shows that only J1108+2530 and J1256+3440 have $\rapo >
15.0$ kpc and $e > 0.7$ and the majority of our sample stars pass the Galactic
center at $\rperi=4.16$. Also, 66$\%$ of our sample stars travel at least 7 kpc
above or below the Galactic plane. 

An additional tool that can be used for this analysis is the Lindblad diagram.
By plotting the total orbital energy vs. angular momentum in the vertical
direction, one can assess the accretion origin of the sample stars. \citet{2014ApJ...788..180C} explored
kinematics, integrals of motion, and orbital properties of 323 VMP stars, to
establish a method to assign membership to the inner- and outer-halo
populations. In this context, stars with total energy $> -0.9$ km$^{2}$ s$^{-2}$
and $\rapo~>15$ kpc can be considered as outer-halo stars. Otherwise, stars can
be considered as members of the inner-halo population.

\begin{figure}[t!]
\epsscale{1.2}
\plotone{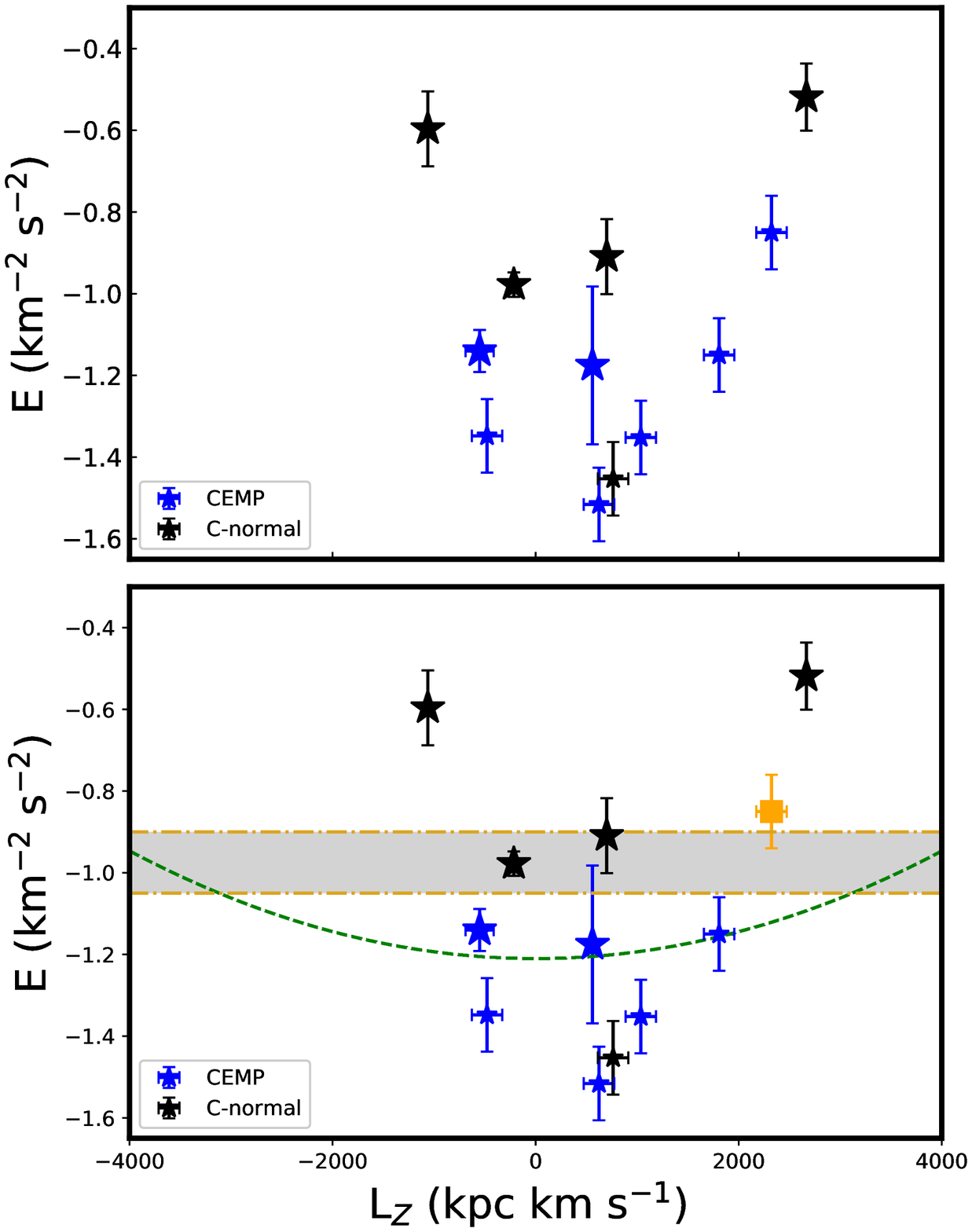} 
\caption{Lindblad diagram for program stars and data taken from Paper I. Top
panel shows the distribution of this sample based on \abund{C}{Fe} abundance
ratios, carbon normal (black stars) and CEMP stars (red stars). Bottom panel
shows the distribution of this sample based on \citet{2014ApJ...788..180C}
criterion. CEMP-no stars are shown as blue filled star, CEMP-r/s as orange
filled square, and C-normal stars as black filled star. The green dashed curve
represents the locus of the points that possess constant apo-Galactic radius,
rapo = 15 kpc. The light-gray shaded area encloses the transition zone
energies.}
\label{fig:Lindblad}
\end{figure}

Figure \ref{fig:Lindblad} shows the Lindblad diagram (top panel) and the
\citet{2014ApJ...788..180C} criterion (bottom panel) for the sample stars,
compared to stars taken from Paper I. CEMP-no stars are shown as blue filled
stars, CEMP-r/s star as orange filled square, and C-normal stars as black filled
stars. The green dashed curve represents the locus of the points that possess
constant apo-Galactic radius, $\rapo = 15$ kpc. The light-gray shaded area
encloses the transition zone energies. It is possible to see that J1630+0953
and J2216+0246, within the error bars, are likely to have inner-halo kinematics. 

Many numerical cosmological simulations suggest that the main origin of the
Milky Way inner and outer-halo stars are massive and low-mass subgalactic fragments,
respectively \citep{2009ApJ...702.1058Z,2011MNRAS.416.2802F,2012MNRAS.420.2245M,
2012AAS...21922206B,  2013MNRAS.432.3391T, 2014MNRAS.439.3128T}. In general, We
can understand the origin of our stars by examining a combination of their
orbital parameters and integrals of motion. The derived orbital parameters and
the calculated total energy of J1630+0953 and J2216+0246 suggest that they
probably belong to the inner-halo population. \textbf{However, their metallicity
and C-enhancement indicate that they may have formed not in situ but in small
mass subgalactic fragments that were accreted very early on and contributed to
the old central regions of the halo system
\citep[e.g.,][]{2018MNRAS.473.1656T}.}
  
\section{Conclusions} \label{sec:conclusion}
 
In this paper, we analyzed six metal-poor stars chosen from the LAMOST database
and followed-up, for the first time, with high-resolution observations using
Lick/APF.  We have presented stellar parameters and full detail chemical
abundances (23 individual elements from C to Eu), for these stars. Our analysis
shows no significant differences compared with the general abundance trends as
reported in previous studies of metal-poor stars. In particular, J1630+0953 and
J2216+0246 present a different behavior (\abund{C}{Fe} $\geqslant +0.7$ and
\abund{Ba}{Fe} $< 0.0$) and are classified as new members of the CEMP-no class. 

We have attempted to characterize the formation scenario and the progenitors of
our CEMP-no stars. Our program stars are located above the solar D$_{trans}$
values and the Forbidden Zone, suggesting that they may have likely formed from
a gas cloud exhibiting fine-structure cooling process. Furthermore, we have
compared $10^{4}$ generated sets of the determined chemical abundances of the
light-element abundances of J1630+0953 and J2216+0246 with predicted yields from
nonrotating massive-star models. About $\sim ~94\%$ of the models predicted that
the mass and explosion energy of the J1630+0953 progenitor, could be in the
21-25 M$_{\odot}$ mass range and $0.3 \times 10^{51}$ erg, respectively, and
only about $\sim ~6\%$ with mass of 50 M$_{\odot}$ and explosion energy of $10.0
\times 10^{51}$ erg.  In about $\sim ~85\%$ of the models predicted 13-16
M$_{\odot}$ mass range and explosion energy  of $1.5$-$1.8 \times 10^{51}$ erg
for the progenitor of J2216+0246, while the remaining models ($\sim ~15\%$)
predicted 10.6-29.5 M$_{\odot}$ mass range and explosion energy $0.6$-$5
\times 10^{51}$ erg. In general, our comparison suggested that massive
stellar progenitors shall be the pollutant source of their birth cloud, and
these pollutants then acted as cooling agents. Our result is consistent with
recent conclusions given by \citet{2018ApJ...857...46I}, which suggest
possible progenitors in the 15-25 M$_{\odot}$ mass range. \textbf{This peak
($\sim$ 20 M$_{\odot}$) may reflect the PopIII initial mass function.
However, this brings the possibility as to whether more massive SN might be
more energetic and therefore destroy their host halo and not allow for EMP
star formation afterwards}.

We have further investigated the kinematics and dynamics of the sample stars,
based on Gaia DR2 data. Our results show that all stars are members of the
inner-halo population. Nevertheless, the deficiency in iron and enhancement in
carbon abundances of J1630+0953 and J2216+0246 strongly suggest that these stars
were born in low-mass sub-galactic systems and later accreted during the initial
phases of Galaxy assembly and contributed to the old stellar populations of the
inner halo.

\section*{Acknowledgement} We thank the anonymous referee for a prompt and
constructive report.  M. K. M. thanks The World Academy of Sciences and the
Chinese Academy of Sciences for the CAS-TWAS fellowship.  M. K. M. thanks Ian
Roederer and Tilman Hartwig for valuable discussions and helpful comments on
earlier versions of the manuscript.  M. K. M. thanks Monika Adam\`{o}w for her
help with using pyMOOGi code.  This study is supported by the National Natural
Sciences Foundation of China under grant No. $11890694$ and $11573032$.  V.M. P.
acknowledges partial support for this work from grant PHY 1430152; Physics
Frontier Center/JINA Center for the Evolution of the Elements (JINA-CEE),
awarded by the US National Science Foundation (NSF). 

\software{IRAF~\citep{1986SPIE..627..733T, 1993ASPC...52..173T}, TAME~
\citep{2015ascl.soft03003K},
MOOG~\citep{1973PhDT.......180S,2011AJ....141..175S, 2017AAS...23021607A},
STARFIT~\citep{2010ApJ...724..341H} ,R-project~ \citep{rproject},
NumPy~\citep{2011CSE....13b..22V}, SciPy~\citep{SciPy2},
Matplotlib~\citep{Hunter:2007}, Astropy~\citep{2013A&A...558A..33A,
2018AJ....156..123A}}.

\bibliography{APF_2nd}
\bibliographystyle{aasjournal}

\begin{deluxetable*}{lllccrrcr}
\tablenum{1}
\tablecolumns{9}
\tabletypesize{\scriptsize}
\tablewidth{0pc}
\tablecaption{Log of the Lick/APF Observations \label{tab:obs}}
\tablehead{
\colhead{}&\colhead{ID}& \colhead{Date}& \colhead{RA}& \colhead{DEC}& \colhead{$r$}& \colhead{Exptime}&
\colhead{S/N}& \colhead{$V_{r}$}\\
\colhead{}& \colhead{}& \colhead{}&  &\colhead{}&\colhead{(mag)}& \colhead{(s)}&
\colhead{(pixel$^{-1})$}& \colhead{(km s$^{-1}$)}}
\startdata
1 &J0326+0202&   18 Nov 2015 & 03 26 53.88 & +02 02 28.1&11.55     & 1800*4&  45&116.2\\  
2 &J1108+2530&   16 Mar 2015 & 11 08 47.18 & +25 30 47.2 &12.16    & 1800*4&  32&$-$114.62\\
3 &J1256+3440&   30 May 2015& 12 56 42.41&  +34 40 58.9 &12.64    & 1800*4&  30&284.06\\
4 &J1413+1727&   24 Jun 2015 & 14 13 15.67 & +17 27 20.8 &11.85    & 1800*4&  42&106.18\\  
5 &J1630+0953&   23 Jun 2015 & 16 30 35.82 & +09 53 17.0 &12.98    & 1800*4&  30&57.82\\  
6 &J2216+0246&   24 Jun 2015 & 22 16 35.96 & +02 46 17.0 &12.47    & 1800*4&  32&$-$94.88\\
\hline
\hline
\tableline
\enddata
\tablecomments{{\footnotesize The S/N ratio per pixel was measured using IRAF at $\lambda \sim 4500$\,{\AA}\,.\\ }}
\end{deluxetable*}

\begin{deluxetable*}{lllrrrrrrrrr}
\tablenum{2}
\tablecaption{Equivalent Widths of Our Sample.\label{tab:EWs}}
\tabletypesize{\scriptsize}
\tablecolumns{22}
\tablewidth{0pt}
\tablehead{\colhead{$\lambda$}& \colhead{ Species  }& \colhead{$ \chi$ }& \colhead{\ensuremath{\log\,gf} }&\colhead{J0326+0202}& \colhead{J1108+2530}& \colhead{J1256+3440 }& \colhead{J1413+1727} & \colhead{J1630+0953}   & \colhead{J2216+0246}         \\
                                               \colhead{({\AA}\,) }  &      &   \colhead{eV }  & &  \colhead{~~~~m{\AA}\,} &  \colhead{~~~~~~~~~m{\AA}\,}  &  \colhead{~~~~~~~~~m{\AA}\,}  &  \colhead{~~~~~~~~~m{\AA}\,}  &  \colhead{~~~~~~~~~m{\AA}\,}  &  \colhead{~~~~~~~~~m{\AA}\,}  }
\startdata
4312      &    C(CH)     &  ---    &    ---   &   syn &   syn  &   syn &  syn &  syn  &   syn   \\
4323      &    C(CH)     &  ---    &    ---   &   syn &   syn  &   syn &  syn &  syn  &   syn   \\
4214      &    N(CN)     &  ---    &    ---   &   syn &   syn  &   syn &  syn &  syn  &   syn   \\
%6970      &    N(CN)     &  ---    &    ---   &   syn &   syn  &   syn &  syn &  syn  &   syn   \\
5889.95 & \ion{Na}{1} &  0.00 &  0.10 & 99.5  &132.5 &146.6 &  158.0 &144.0 &142.3   \\
5895.92 & \ion{Na}{1} &  0.00 & $-$0.20 &122.0 &115.2 & 58.3 &  125.0 &136.3 &110.4   \\
4057.50 & \ion{Mg}{1} &  4.35 & $-$0.89 & 45.9 & 31.6 & 24.0 &--- &--- & 26.0   \\
4167.27 & \ion{Mg}{1} &  4.35 & $-$0.71 & 32.7 & 31.5 & 45.4 &--- & 33.3 & 59.0   \\
4571.10 & \ion{Mg}{1} &  0.00 & $-$5.69 & 22.4 & 46.3 &--- &34.3 & 55.8 & 68.6   \\
4702.99 & \ion{Mg}{1} &  4.33 & $-$0.44 & 44.5 & 57.2 & 24.1 &  53.7 & 73.4 & 76.7   \\
5172.68 & \ion{Mg}{1} &  2.71 & $-$0.45 &--- &--- &--- &168.1&--- &---   \\
5183.60 & \ion{Mg}{1} &  2.72 & $-$0.24 &--- &--- &--- &178.8 &--- &---   \\
5528.40 & \ion{Mg}{1} &  4.35 & $-$0.50 & 42.7 & 60.2 & 41.3 &  48.3 & 58.9 & 88.4   \\
4283.01 & \ion{Ca}{1} &  1.89 & $-$0.22 &--- & 28.3 & 27.0 &  40.6 & 34.6 & 62.2   \\
4318.65 & \ion{Ca}{1} &  1.89 & $-$0.21 &--- & 38.7 & 40.2 &39.1 & 47.9 & 50.4   \\
4425.44 & \ion{Ca}{1} &  1.88 & $-$0.36 & 14.2 & 26.6 & 32.9 &35.7& 26.9 & 28.5   \\
4454.78 & \ion{Ca}{1} &  1.90 &  0.26 & 50.4 & 51.6 & 23.2 &60.7 & 42.5 & 70.0   \\
4455.89 & \ion{Ca}{1} &  1.90 & $-$0.53 & 17.1 & 30.3 &--- &--- & 15.4 & 27.3   \\
5261.71 & \ion{Ca}{1} &  2.52 & $-$0.58 &  2.7 &--- &--- &  9.1 &  8.5 &---   \\
5265.56 & \ion{Ca}{1} &  2.52 & $-$0.11 &  7.5 & 21.1 &--- &17.1 & 17.8 & 17.5   \\
5512.98 & \ion{Ca}{1} &  2.93 & $-$0.45 &  2.4 &  3.4 &--- &  1.8 &--- &---   \\
5581.97 & \ion{Ca}{1} &  2.52 & $-$0.56 &  2.0 &  9.0 &--- &5.4 & 10.8 & 16.1   \\
5588.76 & \ion{Ca}{1} &  2.53 &  0.36 &--- & 40.2 &--- &  32.5 & 31.8 & 50.7   \\
5590.12 & \ion{Ca}{1} &  2.52 & $-$0.57 &  3.8 &  7.9 &--- &  7.5 &  2.3 & 16.5   \\
5594.47 & \ion{Ca}{1} &  2.52 &  0.10 & 15.0 & 23.5 &--- &  1.1 & 29.4 & 54.5   \\
\hline
\hline
\enddata
\tablecomments{{\footnotesize(This table is available in machine-readable form.)}}
\end{deluxetable*}

\begin{deluxetable*}{llllllllllllllllllrrrrr}
\tablenum{3}
\tablecolumns{22}
\tabletypesize{\scriptsize}
\tablewidth{0pc}
\tablecaption{Stellar Parameters of the Program Stars.\label{tab:stellar-param}}
\tablehead{
\colhead{}&\multicolumn{4}{c}{Lick/APF (corrected \& adopted)}&&\multicolumn{4}{r}{Lick/APF (spectroscopic)}&&\multicolumn{2}{c}{Photometry}&&&\multicolumn{2}{l}{~Gaia DR2}&&\\
\cline{2-5}\cline{7-10}\cline{12-13}\cline{16-17}\cline{20-21}\\
\colhead{ID}& \colhead{\Teff }& \colhead{\ensuremath{\log\,g}}& \colhead{[Fe/H]}& \colhead{$\xi$} &
\colhead{}& \colhead{\Teff}& \colhead{\ensuremath{\log\,g}}& \colhead{[Fe/H]}&  \colhead{$\xi$}  &  & \colhead{\Teff (V-J)} &\colhead{\Teff (V-K)}&&&&  \colhead{\ensuremath{\log\,g}}\\
\colhead{}& \colhead{(K)}& \colhead{(cgs)}& \colhead{}& \colhead{(kms$^{-1}$)} &
\colhead{}& \colhead{(K)}& \colhead{(cgs)}& &\colhead{(kms$^{-1}$)}&\colhead{}& \colhead{(K)}& \colhead{(K)}&&&& \colhead{~(cgs)} &&& \colhead{{}}}
\startdata
J0326+0202&  5080  &      2.03   &    $-$3.11&  ~~2.01&&		4900 &      1.59  &     $-$3.26 &  ~~1.92&&		~~~~5083& ~~~~5187&  &&&~~~~~2.06    \\
J1108+2530&  5003  &      2.05   &    $-$2.72&  ~~1.08&&		4815 &      1.52  &     $-$2.89 &  ~~1.09&&		~~~~5136& ~~~~5141 &  &&&~~~~~2.15    \\
J1256+3440&  5215	&	1.74   &	$-$2.76&  ~~2.37&&		5050 &      1.36  &     $-$2.89 &  ~~2.30&&		~~~~5074& ~~~~5142 &  &&&~~~~~1.87  \\
J1413+1727&  4914	&	1.25	  &	$-$3.00&	~~1.92&&		4716&	 0.72	 &	$-$3.20 & 	 ~~2.02&&                 ~~~~4709& ~~~~4704&  &&&~~~~~1.23    \\ 
J1630+0953&  4783	&	1.02	  &	$-$2.94&	~~2.14&&		4570&	 0.40	 &	$-$3.04 & 	 ~~2.28&&                 ~~~~4693& ~~~~4723 &  &&&~~~~~1.30    \\ 
J2216+0246&  4945	&	1.53   &	$-$2.52&	~~1.68&&		4750	&	 1.05  &	$-$2.67 &	 ~~1.67&&		~~~~4899& ~~~~4936 &  & &&~~~~~1.87    \\
\hline
\hline
\tableline
\enddata
\end{deluxetable*}

\begin{deluxetable*}{lrrcrcrrcrcrrcrcrrcrcc}
\tablecolumns{22}
\tablenum{4}
\tabletypesize{\scriptsize}
\tablewidth{0pt}
\tablecaption{LTE Abundances of Individual Elements for the Program Stars \label{tab:abund}}
\tablehead{
\colhead{}& \multicolumn{4}{c}{J0326+0202}& \colhead{}&
            \multicolumn{4}{c}{J1108+2530}& \colhead{}&
            \multicolumn{4}{c}{J1256$+$3440}& \colhead{}\\
            \cline{2-5}\cline{7-10}\cline{12-15}\\
\colhead{}& \colhead{log\,$\epsilon$(X)}& \colhead{[X/Fe]}& \colhead{$\sigma$}& \colhead{$N$}& \colhead{}&
            \colhead{log\,$\epsilon$(X)}& \colhead{[X/Fe]}& \colhead{$\sigma$}& \colhead{$N$}& \colhead{}&
	    \colhead{log\,$\epsilon$(X)}& \colhead{[X/Fe]}& \colhead{$\sigma$}& \colhead{$N$}& \colhead{}&}
\startdata
%\ion{Li}{1}& ...&...&...&...&&...&...&...&...&&...&...&...&... \\
C(CH)             &  5.56&0.24&0.04&2&& 5.99&0.28& 0.05& 2&& 6.04&0.37& 0.05& 2   \\
~~(CH)$_{corr}$             &...&0.02&...&...&&  ...&0.04&...&...&&...&0.24&...&... & \\
~~(CH)$_{natal}$             &...&0.26&...&...&&  ...&0.32&...&...&&...&0.61&...&... & \\
N(CN)             &...&...&...&...&&  ...&...&...&...&& ...&...&...&... \\
\ion{Na}{1}    &     3.43    &     0.30    &     0.18    &       3&& 3.54   &    0.02   &    0.03   &      2&&3.64&0.19& 0.14&2 \\
\ion{Mg}{1}    &     5.11    &     0.62    &     0.16    &       5&&5.26   &    0.38   &    0.16   &      6&&  5.09 &0.28&0.12& 4\\
\ion{Ca}{1}    &     3.52    &     0.29    &     0.14    &      18&&3.81   &    0.19   &    0.13   &     20&& 3.77    &   0.19    &   0.11     &  11 \\
\ion{Sc}{2}    &     0.26    &     0.22    &     0.13    &       6&&0.36   &   $-$0.07   &    0.11   &     13&&0.34  &    $-$0.05   &    0.18     &   9\\
\ion{Ti}{1}    &     2.24    &     0.40    &     0.09    &      15&& 2.32   &    0.09   &    0.12   &     19&& 2.51  &     0.32  &     0.19   &    13 \\
\ion{Ti}{2}    &     2.06    &     0.22    &     0.11    &      28&&2.06   &   $-$0.17   &    0.17   &     33&&2.25     & 0.06      &  0.15     &  27 \\
\ion{V}{1}    &     0.71    &  $-$0.11    &     0.12    &       1&&0.67   &   $-$0.54   &    0.11   &      1&& 1.01     & $-$0.16     &  0.10    &   1 \\
\ion{Cr}{1}    &     2.51    &  $-$0.02    &     0.08    &       7&& 2.49   &   $-$0.43   &    0.16   &     12&&2.63     & $-$0.25     &  0.16     &   7 \\
\ion{Mn}{1}    &     2.07    & $-$0.25    &     0.12    &       2&& 2.23   &   $-$0.48   &    0.11   &      3&& 2.32     & $-$0.35     &  0.12     &   1\\
\ion{Fe}{1}    &     4.39    &  0.00    &     0.16    &     118&& 4.78   &   $-$0.00   &    0.12   &    131&&4.74     & 0.00   &  0.16     & 123 \\
\ion{Fe}{2}    &     4.38    &  $-$0.01    &     0.18    &      13&&    4.77   &   $-$0.01   &    0.12   &     20&&4.74     & 0.00    &  0.15     &  14 \\
\ion{Co}{1}    &     2.41    &     0.53    &     0.12    &       2&& 2.14   &   $-$0.13   &    0.15   &      3&&2.38     & 0.15     &  0.10     &   1\\
\ion{Ni}{1}    &     3.38    &     0.27    &     0.18    &       7&&3.46   &   $-$0.04   &    0.17   &     12&&3.70     &  0.24     &  0.19     &   6\\
\ion{Zn}{1}    &     1.77    &     0.32    &     0.13    &       1&&1.75   &   $-$0.09   &    0.17   &      2&&2.12     & 0.32     &  0.11     &   1\\
\ion{Sr}{2}    &  $-$0.65    &  $-$0.41    &     0.11    &       1&&$-$0.06   &   $-$0.21   &    0.19   &      1&&  $-$0.46&$-$0.57&0.11   &      1 \\
\ion{Y}{2}     &$-$1.00& $-$0.10&                    0.08&      1  &&$-$1.14   &   $-$0.63   &    0.11   &      3&&$-$1.05     & $-$0.50    &  0.12     &   1 \\
\ion{Zr}{2}    &  $-$0.36    &     0.17    &     0.15   &       1&&$-$0.78   &   $-$0.64   &    0.14   &      4                                        &&...&...&...&... \\
\ion{Ba}{2}   &  $-$1.48    &  $-$0.55    &     0.18    &       2&&$-$0.76   &   $-$0.22   &    0.13   &      2&&$-$1.25     & $-$0.67     &  0.19     &   3\\
\ion{La}{2}     &...&...&...&...                                               &&$-$1.13   &    0.01   &    0.22   &      2&&$-$1.01     &  0.65     &  0.20     &   2 \\
\ion{Ce}{2}    &  $-$1.88    &     $-$0.35    &     0.13    &       1&&$-$2.10   &   $-$0.10   &    0.20   &      1&& $-$1.57     &  $-$0.39    &  0.22     &   2\\
%\ion{Pr}{2}     &...&...&...&...                                             && $-$1.19   &    0.11   &    0.15   &      2                                         &&...&...&...&... \\
\ion{Nd}{2}        &  $-$1.47    &     0.22    &     0.15    &       1&&...&...&...&...                                          &&$-$1.13     &  0.21    &  0.15     &   2 \\
\ion{Sm}{2}     &...&...&...&...                                             && $-$1.25   &    0.50   &    0.19   &      3   &&$-$1.79     & 0.01     &  0.14     &   1\\
\ion{Eu}{2}       &$<-$2.60&$-$0.01&0.13&       1             && $-$2.35   &   $-$0.15   &    0.11   &      1 &&$<-$2.10&0.14&0.14&1 \\
\hline
\hline
\tableline
\enddata
\end{deluxetable*}

\begin{deluxetable*}{lrrcrcrrcrcrrcrcrrcrcc}
\tablecolumns{22}
\tablenum{4}
\tabletypesize{\scriptsize}
\tablewidth{0pt}
\tablecaption{Continued.}
\tablehead{
\colhead{}& \multicolumn{4}{c}{J1413+1727}& \colhead{}&
            \multicolumn{4}{c}{J1630+0953}& \colhead{}&
            \multicolumn{4}{c}{J2216+0246}& \colhead{}\\
            \cline{2-5}\cline{7-10}\cline{12-15}\\
\colhead{}& \colhead{log\,$\epsilon$(X)}& \colhead{[X/Fe]}& \colhead{$\sigma$}& \colhead{$N$}& \colhead{}&
            \colhead{log\,$\epsilon$(X)}& \colhead{[X/Fe]}& \colhead{$\sigma$}& \colhead{$N$}& \colhead{}&
	    \colhead{log\,$\epsilon$(X)}& \colhead{[X/Fe]}& \colhead{$\sigma$}& \colhead{$N$}& \colhead{}&}
\startdata
%\ion{Li}{1}& ...&...&...&...&&...&...&...&...&&...&...&...&... \\
C(CH)             &  5.14&$-$0.29&0.07&2&& 6.13&0.64& 0.05& 2&& 6.19&0.28& 0.09& 2   \\
~~(CH)$_{corr}$             &...&0.65&...&...&&  ...&0.62&...&...&&...&0.42&...&... & \\
~~(CH)$_{natal}$             &...&0.36&...&...&&  ...&1.26&...&...&&...&0.70&...&... & \\
N(CN)             &...&...&...&...&&  5.77&0.88&0.17&1&& ...&...&...&... \\
\ion{Na}{1}    &      3.58     &    0.34      &     0.13   &       3&&       3.50    &      0.20    &      0.11    &        2&&      3.76    &      0.04    &      0.15    &        3\\
\ion{Mg}{1}    &      4.96    &    0.36   &     0.16    &       2&&5.22    &      0.56    &      0.12    &        5&&     5.50    &      0.42    &      0.12    &        6\\
\ion{Ca}{1}    &     3.69    &    0.35   &     0.18    &      11&&      3.61    &      0.21    &      0.19    &       18&&      4.08    &      0.26    &      0.12    &       17 \\
\ion{Sc}{2}    &    0.30    &    0.15   &     0.17    &       7&&      0.20    &     $-$0.01    &      0.13    &       10&&      0.59    &     $-$0.05    &      0.18    &        9\\
\ion{Ti}{1}    &     2.27    &    0.32    &     0.11    &       9&&      2.22    &      0.21    &      0.17    &       14&&       2.66    &      0.23    &      0.12    &       14 \\
\ion{Ti}{2}    &      2.25    &    0.30    &     0.12    &      14&&      2.23    &      0.22    &      0.13    &       28&&      2.34    &     $-$0.09    &      0.13    &       32 \\
\ion{V}{1}    &       0.88   &    $-$0.05    &     0.16    &       1&&      1.10    &      0.11    &      0.15    &        1&&       0.89    &     $-$0.53    &      0.15    &        1\\
\ion{Cr}{1}    &     2.76   &    0.12   &     0.11    &       6&&      2.67    &     $-$0.01    &      0.16    &        8&&      3.01    &     $-$0.11    &      0.17    &        7\\
\ion{Mn}{1}    &    2.33    &     $-$0.10   &     0.10    &       1&&      2.23    &     $-$0.26    &      0.16    &        1&&      2.47    &     $-$0.44    &      0.13    &        2\\
\ion{Fe}{1}    &         4.50    &     0.00    &     0.19    &     119&&      4.56    &     $-$0.00    &      0.15    &      127&&      4.97    &     $-$0.01    &      0.15    &      139 \\
\ion{Fe}{2}    &       4.50    &     0.00    &     0.17    &      11&&      4.56    &     $-$0.00    &      0.17    &       12&&      4.98    &     $-$0.01    &      0.19    &       14 \\
\ion{Co}{1}    &    2.22    &    0.23   &     0.11    &       2&&      2.61    &      0.56    &      0.12    &        1&&      2.42    &     $-$0.05    &      0.19    &        2\\
\ion{Ni}{1}    &      3.20    &     -0.02    &     0.17    &       7&&      3.34    &      0.06    &      0.18    &        9&&      3.63    &     $-$0.07    &      0.17    &       10\\
\ion{Zn}{1}    &   1.88    &     0.32    &     0.13    &       1&&      2.04    &      0.42    &      0.13    &        1&&      2.25    &      0.21    &      0.19    &        1\\
\ion{Sr}{2}    &$-$0.06&$-$0.07&0.20   &      1&&     $-$0.73   &     $-$0.66    &      0.12    &        1&&     $-$0.29    &     $-$0.64    &      0.17    &        1 \\
\ion{Y}{2}     & $-$1.50    &    $-$0.38    &     0.19    &       1&&     $-$1.25    &     $-$0.52    &      0.14    &        1&&     $-$0.44    &     $-$0.13    &      0.13    &        2 \\
\ion{Zr}{2}    &  ...&...&...&...                                                 &&     $-$0.32    &      0.04    &      0.16    &        1&&     $-$0.26    &     $-$0.32    &      0.16    &        1\\
\ion{Ba}{2}   &  $-$1.09    &    $-$0.27    &     0.19    &       3&&     $-$0.95    &     $-$0.19    &      0.19    &        3&&     $-$0.43    &      $-$0.09    &      0.22    &        3\\
\ion{La}{2}     &...&...&...&...                                                 &&     $-$1.99    &     $-$0.15    &      0.14    &        1&&     $-$1.04    &      0.39    &      0.16    &        2\\
\ion{Ce}{2}    &   $-$1.62    &     0.13    &     0.20    &       3&&     $-$1.40    &      $-$0.04    &      0.19    &        1&&     $-$0.97    &      $-$0.03   &      0.20    &        1\\
%\ion{Pr}{2}     &  $-$1.27    &     1.34    &     0.17    &       1&&     $-$1.70    &      0.52    &      0.00    &        1&&     $-$1.70    &      0.11    &      0.00    &        1\\
\ion{Nd}{2}     &   $-$1.41    &     0.50    &     0.19    &       9&&     $-$1.45    &      0.07    &      0.28    &        2&&     $-$1.58    &      $-$0.48    &      0.17    &        4 \\
\ion{Sm}{2}     & $-$2.00    &     0.37    &     0.26    &       2&&      $-$1.81    &      0.17    &      0.15    &        1&&...&...&...&...\\
\ion{Eu}{2}     &$<-$2.35&0.13&0.16&1  &&$<-$2.35&0.07&0.15&1&&$<-$2.00&0.00&0.19&1 \\
\hline
\hline
\tableline
\enddata
\tablecomments{{N refers to the number of lines adopted for determination of the elemental abundances.}}
\end{deluxetable*}

\begin{deluxetable*}{lrrrrrr} 
\tablenum{5}
\tablecolumns{3} 
\tabletypesize{\scriptsize}
\tablewidth{0pt}
\tablecaption{Estimated Abundance Uncertainties in the Element Abundance Ratios [X/Fe] for J1630+0953. The Other Sample Stars Yield Quite Similar Results.\label{tab:errors}} 
\tablehead{\colhead{Element}&\colhead{Ion}&\colhead{Random}& \colhead{$\Delta$\mbox{T$_{\rm eff}$}}&\colhead{$\Delta\log g$}& \colhead{$\Delta v_{micr}$}&\colhead{Root Mean}\\ 
\colhead{}&\colhead{}&\colhead{error}&\colhead{+100\,K}& \colhead{$+$0.3\,dex}&\colhead{+0.3\,km\,s$^{-1}$}&\colhead{Square}} 
\startdata 
C &CH& 0.05& 0.19 &  0.11 &    0.00 & 0.23 \\ 
N &NH& 0.17& 0.22 &  0.11 &    0.00 & 0.30 \\ 
Na& 1& 0.11& 0.09 &$-$0.01& $-$0.02 & 0.14 \\ 
Mg& 1& 0.12& 0.09 &$-$0.06&    0.06 & 0.17 \\ 
Ca& 1& 0.19& 0.08 &$-$0.01& $-$0.03 & 0.21 \\ 
Sc& 2& 0.13& 0.07 &  0.11 & $-$0.02 & 0.19 \\ 
Ti& 1& 0.17& 0.10 &$-$0.01& $-$0.01 & 0.19 \\ 
Ti& 2& 0.13& 0.04 &  0.08 & $-$0.12 & 0.19 \\ 
V& 2& 0.15& 0.05 &  0.10 &    0.00 & 0.19 \\ 
Cr& 1& 0.16& 0.12 &$-$0.01& $-$0.03 & 0.20 \\ 
Mn& 1& 0.15& 0.12 &  0.00 & $-$0.01 & 0.19 \\ 
Fe& 1& 0.17& 0.12 &$-$0.02& $-$0.08 & 0.18 \\ 
Fe& 2& 0.12& 0.02 &  0.11 & $-$0.01 & 0.16 \\ 
Co& 1& 0.12& 0.12 &  0.00 & $-$0.03 & 0.17 \\ 
Ni& 1& 0.18& 0.15 &$-$0.03& $-$0.01 & 0.20 \\ 
\hline
\hline
\tableline
\enddata
\tablecomments{{\textbf{Random errors represent the standard error of the mean and the last column represent the root mean squares of these errors.} }}
\end{deluxetable*}

\begin{deluxetable*}{lccccrccrcccccccccr}
\tablenum{6}
\tablecaption{Parallaxes, Proper Motions and Distances  \label{tab:kinematics_input}}
\tablewidth{0pc}
\tablehead{
\colhead{Star} &\colhead{\textit{Gaia} DR2 source ID} &\colhead{$\varpi$} &\colhead{error} & &\colhead{pmra} &
\colhead{error} &  &\colhead{pmdec} &\colhead{error} &  &\colhead{Distance} &\colhead{d1} &\colhead{d2} & &
\cline{3-4}\cline{6-7}\cline{9-10}\cline{12-14}\cline{16-17}
 &\colhead{} &\multicolumn{2}{c}{(mas)} &  &\multicolumn{2}{c}{(mas~yr$^{-1}$)} &  &
\multicolumn{2}{c}{(mas~yr$^{-1}$)} &  &\multicolumn{3}{c}{(kpc)}}
\startdata
J0326+0202  &   3268028903151246720  &  0.5421    &  0.0392    & &    22.759    &  0.074   & &  $-$14.128  &  0.062    & &  1.727  &  1.617  &  1.853\\
J1108+2530  &   3995801795674031616  &  0.4887    &  0.0423    & &      0.553    &  0.089   & &  $-$38.926  &  0.062    & &  1.893  &  1.753  &  2.057\\
J1256+3440  &   1515900293283319168  &  0.1945    &  0.0469    & &     4.068    &  0.058    & &  $-$13.785  &  0.048    & &  3.601  &  3.134  &  4.194\\
J1413+1727  &   1233499867083517952  &  0.1525    &  0.0488    & &$-$6.039    &  0.089    & &  $-$10.405  &  0.095    & &  4.240  &  3.590  &  5.100\\
J1630+0953  &   4458577516730343424  &  0.1170    &  0.0261    & &$-$8.735    &  0.033    & &    $-$8.939  &  0.023    & &  6.183  &  5.354  &  7.267\\
J2216+0246  &   2682719929806900096  &  0.2893    &  0.0530    & &     8.700    &  0.090    & &  $-$11.429  &  0.100    & &  2.899  &  2.521  &  3.391\\
\hline
\hline
\tableline
\enddata
\tablecomments{The d1 and d2 columns indicate the 16th percentile and 84th percentile confidence intervals.}
\end{deluxetable*}

\begin{deluxetable*}{lDDDcDDDcDDDrr}
\tablenum{7}
\tablecaption{Positions and Galactic Space-velocity Components. \label{tab:UVW}}
\tablewidth{0pt}
\tabletypesize{\scriptsize}
\tablehead{
\colhead{Star} &\twocolhead{X} &\twocolhead{Y} &\twocolhead{Z} &\colhead{} &\twocolhead{U} &\twocolhead{V} &\twocolhead{W} 
&\colhead{} &\twocolhead{$V_{R}$} &\twocolhead{$V_{\phi}$}&&\twocolhead{$V_{\perp}$}  \\
\cline{2-7}\cline{9-14}\cline{16-23} \colhead{} &\multicolumn{6}{c}{(kpc)} &\colhead{} &\multicolumn{6}{c}{(km s$^{-1}$)} &\colhead{} 
& &\multicolumn{6}{c}{(km s$^{-1}$)} &\colhead{}  }
\decimals
\startdata
J0326+0202 & 9.47^{+0.08}_{$-$0.08}       &      0.03^{+0.00}_{$-$0.00}       &$-$1.15^{+0.08}_{$-$0.07}       & &$-$115.54^{+2.84}_{$-$2.79}   &$-$201.44^{+13.81}_{$-$13.13}   &$-$22.65^{+3.24}_{$-$3.39}     & &$-$115.55^{+2.84}_{$-$2.79}   & 201.44^{+13.13}_{$-$13.81}   &  117.74^{+4.31}_{$-$4.39}    \\
J1108+2530 & 8.83^{+0.05}_{$-$0.05}       &      0.38^{+0.03}_{$-$0.03}       &      1.77^{+0.13}_{$-$0.13}       & &   183.79^{+9.97}_{$-$10.14}   &$-$287.20^{+24.04}_{$-$23.70}   &$-$120.18^{+1.86}_{$-$1.89}   & &   183.57^{+9.94}_{$-$10.10}   & 287.33^{+23.72}_{$-$24.05}   &   219.41^{+10.11}_{$-$10.27}    \\
J1256+3440 & 8.40^{+0.03}_{$-$0.03}       & $-$0.43^{+0.06}_{$-$0.06}       &      3.59^{+0.46}_{$-$0.47}       & &   182.38^{+23.98}_{$-$24.38} &$-$110.44^{+20.46}_{$-$19.99}   &317.44^{+3.91}_{$-$3.86}        & & 182.48^{+24.02}_{$-$24.40}   & 110.27^{+19.95}_{$-$20.42}   &   366.15^{+24.34}_{$-$24.70}    \\
J1413+1727 & 6.69^{+0.23}_{$-$0.23}       & $-$0.27^{+0.04}_{$-$0.04}       &      3.99^{+0.60}_{$-$0.60}       & &   92.72^{+7.03}_{$-$6.93}       &$-$220.29^{+36.13}_{$-$35.81}   &104.85^{+0.87}_{$-$0.90}        & &    92.875^{+7.08}_{$-$6.98}    & 220.23^{+35.80}_{$-$36.11}   &    140.07^{+7.13}_{$-$7.04}    \\
J1630+0953 & 3.67^{+0.60}_{$-$0.62}       & $-$2.15^{+0.28}_{$-$0.29}       &      3.63^{+0.49}_{$-$0.47}       & &   142.91^{+12.36}_{$-$11.93}  &$-$311.69^{+44.89}_{$-$46.83}   &131.88^{+12.42}_{$-$12.03}    & &  145.78^{+13.76}_{$-$12.88} &  310.35^{+46.32}_{$-$44.49}   &  196.58^{+18.54}_{$-$17.62}   \\
J2216+0246 & 7.31^{+0.12}_{$-$0.12}       & $-$1.96^{+0.26}_{$-$0.26}       &$-$1.93^{+0.25}_{$-$0.25}        & &$-$37.84^{+2.70}_{$-$2.79}     &$-$187.73^{+17.67}_{$-$17.68}   &$-$72.86^{+18.67}_{$-$18.72} & &$-$36.98^{+2.54}_{$-$2.60}    & 187.90^{+17.70}_{$-$17.70}   &  81.71^{+18.84}_{$-$18.90}   \\
\hline
\hline
\tableline
\enddata
\tablecomments{The $-$ and $+$ indicate the 16th percentile and 84th percentiles}
\end{deluxetable*}

\begin{deluxetable*}{lDDDcDDDcDDDrrr}
\tablenum{8}
\tablecaption{Calculated Orbital Parameters, Energies and Angular Momenta. \label{tab:energy}}
\tablewidth{0pt}
\tabletypesize{\scriptsize}
\tablehead{
\colhead{Star} &\twocolhead{$r_{apo}$} &\twocolhead{$r_{peri}$} &\twocolhead{$Z_{max}$} &\colhead{} &\twocolhead{e} &\twocolhead{E} &\twocolhead{$L_{Z}$} 
&\colhead{}  \\
\cline{2-7}\cline{9-14}\cline{16-23} \colhead{} &\multicolumn{6}{c}{(kpc)} &\colhead{} &\multicolumn{6}{c}{~~~~~~~~~~~~~~~~($10^{3}$km$^{2}$ s$^{-2}$)~~~~~~~(kpc km s$^{-1}$)} &\colhead{} 
& & }
\decimals
\startdata
J0326+0202 &10.88^{+0.18}_{$-$0.16}       &0.42^{+0.30}_{$-$0.26}       &1.35^{+0.05}_{$-$0.03}       &&0.93^{+0.05}_{$-$0.05}       &$-$0.98^{+0.03}_{$-$0.03}       &$-$214.66^{+29.32}_{$-$30.21}          \\
J1108+2530 &15.65^{+1.73}_{$-$1.31}       &1.91^{+0.57}_{$-$0.60}       &7.43^{+0.23}_{$-$0.26}       &&0.78^{+0.05}_{$-$0.03}       &$-$0.60^{+0.09}_{$-$0.09}       &$-$1061.80^{+21.31}_{$-$22.01}        \\
J1256+3440 &65.94^{+14.83}_{$-$9.67}     &8.90^{+0.21}_{$-$0.19}       &29.69^{+4.02}_{$-$2.62}     &&0.76^{+0.04}_{$-$0.03}       &$-$0.52^{+0.08}_{$-$0.07}       &     2666.88^{+41.09}_{$-$40.83}          \\
J1413+1727 &8.03^{+0.10}_{$-$0.08}         &2.88^{+0.74}_{$-$0.25}       &7.90^{+0.12}_{$-$0.50}       &&0.47^{+0.03}_{$-$0.09}       &$-$0.91^{+0.09}_{$-$0.08}       &       700.25^{+25.89}_{$-$25.67}          \\
J1630+0953 &7.11^{+2.54}_{$-$0.69}         &4.16^{+0.82}_{$-$1.23}       &7.11^{+2.50}_{$-$0.72}       &&0.30^{+0.11}_{$-$0.04}        &$-$1.18^{+0.02}_{$-$0.02}       &       559.86^{+5.63}_{$-$5.21}            \\
J2216+0246 &8.07^{+0.11}_{$-$0.12}         &0.47^{+0.60}_{$-$0.25}       &3.57^{+2.46}_{$-$0.92}       &&0.89^{+0.06}_{$-$0.13}       &$-$1.14^{+0.05}_{$-$0.04}       &$-$551.44^{+138.48}_{$-$137.99}          \\
\hline
\hline
\tableline
\enddata
\tablecomments{The $-$ and $+$ indicate the 16th percentile and 84th percentiles}
\end{deluxetable*}

\end{document}